\documentclass[aps,prd,reprint,onecolumn,superscriptaddress,nofootinbib,floatfix]{revtex4-1}
\pdfoutput=1
\usepackage{epsfig, subfigure}
\usepackage{setspace}
\usepackage{booktabs, tabularx} 
 
\usepackage{amsmath,bm,bbm}
\usepackage{hyperref}
\usepackage{graphicx}
\usepackage{enumerate}
\usepackage{dcolumn}
\usepackage{bm}
\usepackage{longtable}
\hypersetup{
    pdfnewwindow=true,      
    colorlinks=true,       
    linkcolor=blue,          
    citecolor=blue,        
    filecolor=blue,      
    urlcolor=blue           
}

\def\lsim{\lesssim}  
\def\gsim{\gtrsim}

\newcommand{\beq}{\begin{equation}}  
\newcommand{\eeq}{\end{equation}}
\newcommand{\bea}{\begin{eqnarray}}  
\newcommand{\eea}{\end{eqnarray}}
\newcommand{\epm}{\ensuremath{e^{\pm}\;}}

\newcommand{\omb}{\ensuremath{\Omega_{\rm B} h^{2}}}
\newcommand{\neff}{\ensuremath{{\rm N}_{\rm eff}}}
\newcommand{\Deln}{\ensuremath{\Delta{\rm N}_\nu}}

\newcommand{\mchi}{\ensuremath{m_\chi}}
\newcommand{\yd}{\ensuremath{y_{\rm DP}}}
\def\3he{$^3$He}
\def\4he{$^4$He}
\def\7li{$^7$Li}
\def\Yp{Y$_{\rm P}$}

\newcommand{\ie}{{\it i.e.}}
\newcommand{\eg}{{\it e.g.}}
\newcommand{\etal}{{\it et al.}}

\begin{document}

\title{\mbox{\hspace{-0.5cm}BBN And The CMB Constrain Neutrino Coupled Light WIMPs}}

\author{Kenneth M. Nollett}
 \email{nollett@mailbox.sc.edu}
 \affiliation{\mbox{Department of Physics and Astronomy, University of South Carolina, 712 Main St., Columbia, SC 29208, USA}}
 \affiliation{\mbox{Department of Physics and Astronomy, Ohio University, Athens, OH~~45701, USA}}
 \affiliation{\mbox{Department of Physics, San Diego State University, 5500 Campanile Drive, San Diego, CA~~92182-1233}}

\author{Gary Steigman}
\email{steigman.1@osu.edu}
\affiliation{\mbox{Center for Cosmology and AstroParticle Physics, Ohio State University,}}
\affiliation{Department of Physics, Ohio State University, 191 W.~Woodruff Ave., Columbus, OH 43210, USA}

\date{\today}

\begin{abstract}

In the presence of a light weakly interacting massive particle, a WIMP with mass $m_{\chi} \lsim 30\,{\rm MeV}$, there are degeneracies among the nature of the WIMP (fermion or boson), its couplings to the standard model particles (to electrons, positrons, and photons, or only to neutrinos), its mass \mchi, and the number of equivalent (additional) neutrinos, \Deln.  These degeneracies cannot be broken by the cosmic microwave background (CMB) constraint on the effective number of neutrinos, \neff.  However, since big bang nucleosynthesis (BBN) is also affected by the presence of a light WIMP and equivalent neutrinos, complementary BBN and CMB constraints can help to break some of these degeneracies.  In a previous paper \cite{kngs1} the combined BBN and Planck \cite{planck} CMB constraints were used to explore the allowed ranges for \mchi, \Deln, and \neff~in the case where the light WIMPs annihilate electromagnetically (EM) to photons and/or \epm pairs.  In this paper the BBN predictions for the primordial abundances of deuterium and \4he (along with \3he and \7li) in the presence of a light WIMP that  only couples (annihilates) to neutrinos (either standard model -- SM -- only or both SM and equivalent) are calculated.   Recent  observational estimates of the relic abundances of D and \4he are used to limit the light WIMP mass, the number of equivalent neutrinos, the effective number of neutrinos, and the present Universe baryon density (\omb).  Allowing for a neutrino coupled light WIMP and \Deln~equivalent neutrinos, the combined BBN and CMB data provide lower limits to the WIMP mass that depend very little on the nature of the WIMP (Majorana or Dirac fermion, real or complex scalar boson), with a best fit \mchi~$\gtrsim 35\,{\rm MeV}$, equivalent to no light WIMP at all.  The analysis here excludes all neutrino coupled WIMPs with masses below a few MeV, with specific limits varying from 4 to 9 MeV depending on the nature of the WIMP.  In the absence of a light WIMP (either EM or neutrino coupled), BBN alone prefers $\Deln = 0.50 \pm 0.23$, favoring neither the absence of equivalent neutrinos (\Deln~= 0), nor the presence of a fully thermalized sterile neutrino (\Deln~= 1).  This result is consistent with the CMB constraint, $\neff = 3.30 \pm 0.27$ \cite{planck}, constraining ``new physics" between BBN and recombination.  Combining the BBN and CMB constraints gives $\Deln = 0.35 \pm 0.16$ and $\neff = 3.40 \pm 0.16$.  As a result, while BBN and the CMB combined require $\Deln \geq 0$ at $\sim 98\,\%$ confidence, they disfavor $\Deln \geq 1$ at $> 99\,\%$ confidence.  Adding the possibility of a neutrino-coupled light WIMP extends the allowed range slightly downward for \Deln~and slightly upward for \neff~simultaneously, while leaving the best-fit values unchanged.

\end{abstract}

\maketitle

\section{Introduction}

While weakly interacting massive particles (WIMPs, $\chi$) present in some extensions of the standard model (SM) of particle physics are
usually very massive, with $m_{\chi}$ in excess of tens or hundreds of GeV, there has been a long and continuing interest in light ($m_{e}
\lsim m_{\chi} \lsim$ tens of MeV) or even very light ($m_{\chi} \lsim m_{e}$) WIMPs
\cite{ktw,serpico,boehm1,boehm2,boehm3,hooper1,boehm4,hooper2,ahn,fayet,hooper3,feng}.  More recently, the cosmic microwave background (CMB) has been used to measure the expansion rate of the universe during the acoustic oscillation era, expressed as an effective number of neutrino species, \neff.  Such measurements have led to exploration of the effect on \neff~of a hypothetical WIMP that is sufficiently light to annihilate after neutrino decoupling and heat either the photons or the SM neutrinos beyond the usual heating from \epm annihilation~\cite{chimera,hoscherrer1,hoscherrer2,boehm2012}.  For the standard models of particle physics and cosmology (\ie, no light WIMPs, no equivalent neutrinos), \neff~measured in the late Universe is $\neff = 3$\,\footnote{If the assumption of instantaneous neutrino decoupling (IND) at high temperature is relaxed, \neff~= 3.046 \cite{dolgov,hannestad,mangano}.}.  In those extensions of the SM containing ``dark radiation'' whose contribution to the early Universe energy density amounts to that of \Deln~``equivalent neutrinos,'' it is generally the case that $\neff > 3$.  In Ref.~\cite{chimera} this canonical result was revisited, demonstrating that in the presence of a light WIMP that annihilates only to SM neutrinos, a measurement of $\neff > 3$ from the CMB temperature anisotropy spectrum can be consistent with \Deln~= 0 (``dark radiation without dark radiation").  It was also shown that in the presence of a sufficiently light WIMP that couples strongly to photons and/or \epm pairs, a CMB measurement of $\neff = 3$ is not inconsistent with the presence of dark radiation ($\Delta{\rm N}_{\nu} > 0$) \cite{chimera}.  In other words, the presence of one or more equivalent neutrinos, including ``sterile neutrinos"\,\footnote{Here, the term sterile neutrinos is reserved for equivalent neutrinos with \Deln~= 1.  For several sterile neutrinos, \Deln~is an integer, $\geq 1$.}, can be consistent with $\neff = 3$.  Depending on its couplings, the late time annihilation of a light WIMP will heat either the relic photons or the relic neutrinos, and this will affect the CMB constraint on the sum of the neutrino (SM and equivalent) masses \cite{chimera}.  It was emphasized in Ref.~\cite{chimera} that in the presence of a light WIMP and/or equivalent neutrinos there are degeneracies among the light WIMP mass and its nature (fermion or boson, as well as its couplings to \epm pairs and photons, or to neutrinos), the number and nature (fermion or boson) of the equivalent neutrinos, and the temperature at which they decouple from the SM particles.  Constraints from the CMB alone are insufficient to break these degeneracies.

However, as already shown by Kolb \etal\,\cite{ktw} and Serpico and Raffelt \cite{serpico}, and more recently by B\oe hm
\etal\,\cite{boehm2013}, the presence of a light WIMP modifies the early Universe energy and entropy densities (more specifically, the relation between the photon and neutrino temperatures), affecting the synthesis of the light nuclides during big bang nucleosynthesis (BBN).  As a result, BBN provides additional constraints on the properties of light WIMPs and equivalent neutrinos that, in combination with the information from later epochs provided by the CMB, can help to break some of these degeneracies.  None of the previous studies of BBN in the presence of a light WIMP allowed for equivalent neutrinos, thereby eliminating some potentially interesting possibilities (\eg, sterile neutrinos).  In our previous paper \cite{kngs1} we explored the consequences for BBN of equivalent neutrinos, allowing for a light WIMP (a Majorana or Dirac fermion, or a real or complex scalar boson) that remains in thermal equilibrium with \epm pairs and photons until a low enough temperature that all the WIMPs have annihilated away (similar to the millicharged, light particle proposed by Dolgov \etal~\cite{millicharged}).  In this paper we turn our attention to light WIMPs that remain in equilibrium with, and annihilate into, either the SM neutrinos or both the SM and equivalent neutrinos (like the ``neutrino secret interaction'' mediators discussed in Refs.~\cite{ng,ahlgren}).  As in Ref.~\cite{kngs1}, BBN is revisited, now allowing for neutrino coupled, light WIMPs and equivalent neutrinos.  The 68.3\% and 95.5\% confidence level regions in the multidimensional parameter spaces of various combinations of the WIMP mass (\mchi), the number of equivalent neutrinos (\Deln), the effective number of neutrinos at recombination (\neff), and the baryon density parameter (\omb), are identified and compared with the independent constraints on \neff~and \omb~from the Planck CMB results \cite{planck}.  In this manner the consistency between the physics and the evolution of the Universe at BBN ($\sim$ first few minutes) and at recombination ($\sim 400$ thousand years later) is tested, and the allowed ranges and best fit values of the parameters are identified.

In \S\,\ref{sec:overview}, two of the three key parameters explored here, the effective number of neutrinos (\neff) and the number of equivalent neutrinos (\Deln) are defined more precisely, and the influence of a light WIMP on the \neff~-- \Deln~relation is explored. \S\,\ref{sec:nureview} focuses on the effects on the key parameters of a light WIMP that annihilates only to neutrinos (with and without restriction to SM neutrinos\footnote{A light WIMP that annihilates only to particles in a ``dark sector,'' not to SM particles (similar to that considered, \eg, in Ref.~\cite{feng2}), is beyond the scope of this paper.}).  Since the publication of our previous paper exploring the BBN and CMB constraints on an electromagnetically coupled light WIMP \cite{kngs1}, there has been a new determination of the primordial abundance of deuterium \cite{cooke} that has a small, but non-negligible, effect on the parameter constraints presented there\,\footnote{A subsequent update to the helium abundance, published late in the preparation of this paper, is discussed briefly in \S\,\ref{sec:summary}.}.  In \S\,\ref{sec:nowimp} this new D abundance is used along with the same primordial helium abundance adopted previously \cite{izotov}, to update the BBN constraints on the baryon density and \Deln.  These slightly revised BBN constraints are combined with those on the baryon density and \neff~from Planck \cite{planck} to update the best joint fit values of \neff, \Deln, and \omb.  Then, in \S\,\ref{sec:bbnnuwimp}, the effects on primordial nucleosynthesis of a neutrino coupled light WIMP are outlined as a function of the WIMP mass for a light WIMP that could be a Majorana or Dirac fermion, as well as a real or complex scalar boson.  In \S\,\ref{sec:joint} the joint BBN and CMB constraints on the key parameters are presented.  In particular, the {\it lower} bounds to the masses of the different kinds of neutrino coupled light WIMPs are presented in Table\,\ref{tab:masstable} and the assumptions leading to these bounds are summarized in \S\,\ref{sec:lower bound}.  Finally, in \S\,\ref{sec:summary} the results presented here, as well as those updated for an electromagnetically coupled light WIMP, are summarized.

\section{Review And Overview}
\label{sec:overview}
 
In preparation for the results to be presented here, it is important to establish various definitions relevant to the CMB constraints, following the notation of Ref.~\cite{chimera}.  The key parameters are the effective number of neutrinos, \neff, and the number of equivalent neutrinos, \Deln.  At late times in the early Universe, long after the \epm pairs and any light WIMPs have annihilated, the only particles contributing to the radiation energy density are the photons ($\gamma$), the three SM neutrinos ($\nu_{e}, \nu_{\mu}, \nu_{\tau}$), and \Deln~equivalent neutrinos ($\xi$).  At these late times, when $T_{\gamma} \rightarrow T_{\gamma 0} \ll \{m_{e}\,,m_{\chi}\}$, and $T_{\gamma 0}$ is still greater than the neutrino masses, the radiation energy density, normalized to the energy density in photons alone, is 
\beq 
\bigg({\rho_{\rm R} \over \rho_{\gamma}}\bigg)_{0} = 1 + \bigg({\rho_{\nu} \over \rho_{\gamma}}\bigg)_{0}\,\bigg[3 + \bigg({\rho_{\xi} \over  \rho_{\nu}}\bigg)_{0}\bigg] = 1 + \bigg({\rho_{\nu} \over \rho_{\gamma}}\bigg)_{0}\,[3 + \Deln]\,.
\eeq 
The ``number of equivalent neutrinos'' appearing here, \Deln, is defined by the late time (but when the neutrinos are still relativistic) ratio of the energy density in one, or more, extremely relativistic (ER) $\xi$ particles (fermions or bosons) to that in one of the SM neutrinos, $\Deln \equiv (\rho_{\xi}/\rho_{\nu})_{0}$.  In the absence of a light WIMP the SM neutrinos decouple at a temperature $T_{\nu d}$, between 2 and 3 MeV, prior to the bulk of the \epm annihilation \cite{enqvist1992}.  When the \epm pairs annihilate, they (mainly) heat the photons relative to the neutrinos.  As a result, although $T_{\nu} = T_{\gamma}$ for $T_{\gamma} \geq T_{\nu d}$, once $T_{\gamma} < T_{\nu d}$, it is the case that $T_{\gamma} > T_{\nu}$.  The canonical, textbook assumption is that the SM neutrinos decouple instantaneously, when the only other thermally populated relativistic particles are the photons and the \epm pairs, and also that at neutrino decoupling the \epm pairs are extremely relativistic ($m_{e}/T_\gamma \rightarrow 0$).  Under these assumptions the neutrino phase space distribution remains that of a relativistic Fermi gas, so that $(\rho_{\nu}/\rho_{\gamma})_{0} = 7/8\,(T_{\nu}/T_{\gamma})_{0}^{4}$, and entropy conservation can be applied to find that, $(T_{\nu}/T_{\gamma})^{3}_{0} = 4/11$.  The above result, where \epm annihilation heats the photons relative to the decoupled SM and equivalent neutrinos, is used to define the ``effective number of
neutrinos,'' \neff:
\beq 
\bigg({\rho_{\rm R} \over \rho_{\gamma}}\bigg)_{0} = 1 + {7 \over 8}\bigg({T_{\nu} \over T_{\gamma}}\bigg)^{4}_{0}\,[3 + \Deln]\equiv 1 + {7 \over 8}\bigg({4 \over 11}\bigg)^{4/3}{\rm N}_{\rm eff}\,, 
\eeq 
where, 
\beq 
{\rm N}_{\rm eff} = \bigg[{11 \over 4}\bigg({T_{\nu} \over T_{\gamma}}\bigg)^{3}_{0}\bigg]^{4/3}[3 + \Deln] = 3\,\bigg[{11 \over 4}\bigg({T_{\nu} \over T_{\gamma}}\bigg)^{3}_{0}\bigg]^{4/3}\bigg[1 + {\Deln \over3}\bigg] = \neff^{0}\,\bigg[1 + {\Deln \over 3}\bigg]\,,
\label{eq:original-neff}
\eeq 
and (denoting the $\Deln = 0$ case with a superscript zero), 
\beq 
\neff^{0} = 3\,\bigg[{11 \over 4}\bigg({T_{\nu} \over T_{\gamma}}\bigg)^{3}_{0}\bigg]^{4/3}\,.  
\eeq 
It is important to keep in mind that \neff~is defined to be a ``late-time'' quantity, characterizing the SM and equivalent neutrino contributions to the relativistic energy density after BBN has ended, and the \epm pairs and any light WIMPs (considered later) have annihilated.  The evolution of the energy density during BBN is {\bf not} described as an evolution of \neff~with time.

Retaining the assumption of instantaneous neutrino decoupling (IND), but relaxing the assumption that the \epm pairs are ER at neutrino
decoupling, 
\beq 
\bigg({T_{\nu} \over T_{\gamma}}\bigg)^{3}_{0} = {4 \over 2g_{s}(T_{\nu d}) - 10.5}\,, 
\eeq 
where $g_{s}(T)$ is defined by the ratio of the total entropy to the entropy in photons alone \cite{kngs1,chimera}.  In the approximation that $m_e/T_{\nu d} = 0 $, $g_{s}(T_{\nu d}) = 10.75$, so that the canonical results, $(T_{\nu}/T_{\gamma})^{3}_{0} = 4/11$ and \neff~= 3, are recovered.  However, at $T_{\nu d} \approx 2\,{\rm MeV}$, the electron/positron entropy is already slightly less than the relativistic limit, so that $2g_{s}(T_{\nu d}) - 10.5 \approx 10.95$, corresponding to $\neff \approx 3.02$.  This is to be compared to the value, $\neff \approx 3.05$, found \cite{mangano} when the IND is relaxed and the neutrino phase space density computed in detail, ignoring any equivalent neutrinos.   As far as we are aware, no detailed calculation of the neutrino phase space distribution \cite{dolgov,hannestad,mangano} has been published that allows for either equivalent neutrinos or light WIMPs, and such a calculation is beyond the scope of the present work.  It is assumed here that $\Deln \neq 0$ will not change the literature solutions very  much, so $\neff \approx 3.05(1 + \Deln/3)$ is adopted when there is no light WIMP (\ie, for $m_{\chi} \gsim 30\,{\rm MeV}$).  However, the IND is used (\eg, in Eq.\,(3)) whenever there is a light WIMP.  In the IND approximation with decoupling at $T_{\nu d} = 2$ MeV, the high WIMP mass limit is $\neff \approx 3.02(1 + \Deln/3)$.  The difference between these approximations ($\approx 0.03 + 0.01\,\Deln$) is very small compared with the errors in the present BBN and CMB constraints on \neff, but may need to be accounted for when higher precision data become available.

\begin{figure}[!t]
\begin{center}
\includegraphics[width=0.5\columnwidth]{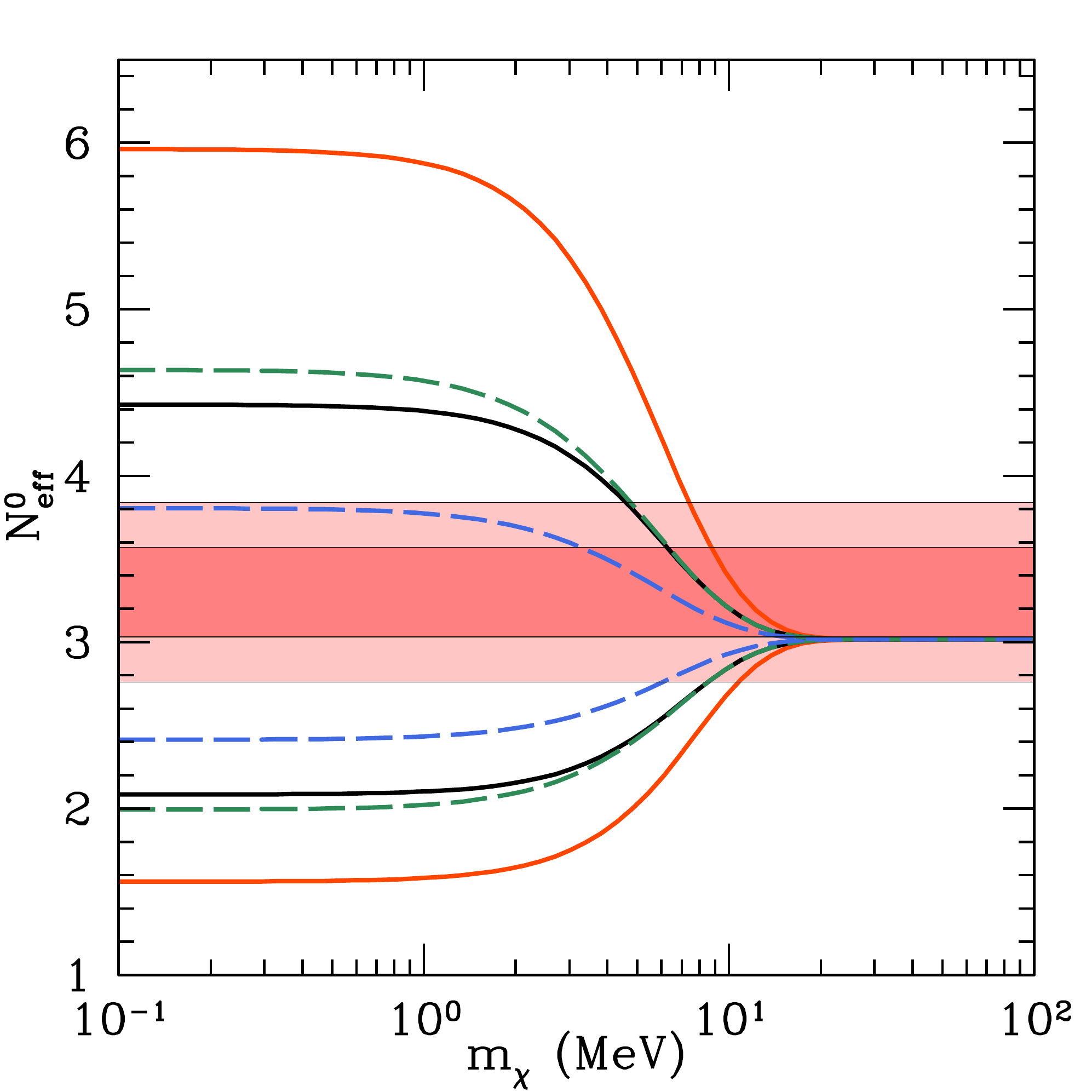}
\caption{(Color online) $\neff^{0}$, the value of \neff~for \Deln~= 0, as a function of the WIMP mass for WIMPs that annihilate to \epm pairs and photons (lower set of curves) and those that annihilate to the SM neutrinos (upper set of curves).  For the top set of curves (neutrino coupled), from top to bottom, the solid (red) curve is for a Dirac WIMP, the dashed (green) curve is for a complex scalar, the solid (black) curve is for a Majorana WIMP, and the dashed (blue) curve is for a real scalar.  The order of the curves is reversed for the lower set (EM coupled).  The horizontal (red/pink) bands are the Planck CMB 68.3\% and 95.5\% ranges for N$_{\rm eff}$.}
\label{fig:neff0vsmall}
\end{center}
\end{figure}

In the presence of a light Majorana WIMP that annihilates to \epm pairs and/or photons (i.e., is EM coupled), the photons are further heated relative to the already decoupled SM and equivalent neutrinos, and $\neff^{0}$ is a function of the WIMP mass (see, \eg, \cite{kngs1,chimera} for details).  In this case
\beq
\neff^{0} = 3\left[\frac{11}{4}\bigg({T_{\nu} \over T_{\gamma}}\bigg)^{3}_{0}\right]^{4/3} 
= 3\bigg[{11 \over 10.95 + 3.5\,\phi_{\chi d}}\bigg]^{4/3}\,,
\label{eq:neff0-nowimp}
\eeq
where $\phi_{\chi d}$ is the ratio of the entropy carried by the light WIMP at $T_{\nu d}$ to the entropy it would have contributed were it massless.  The $\neff^{0} - m_{\chi}$ relation is shown for several possibilities of EM coupled WIMPs by the lower set of curves in Figure \ref{fig:neff0vsmall}.  In the high mass limit, $\neff^{0} \rightarrow 3.02$.  In the limit of very low WIMP mass, $\neff^{0} < 3.02$, ranging from $\neff^{0} \approx 1.6$ to $\neff^{0} \approx 2.4$, with the value depending on the statistics obeyed by the WIMP.

\begin{figure}[!t]
\begin{center}
\includegraphics[width=0.45\columnwidth]{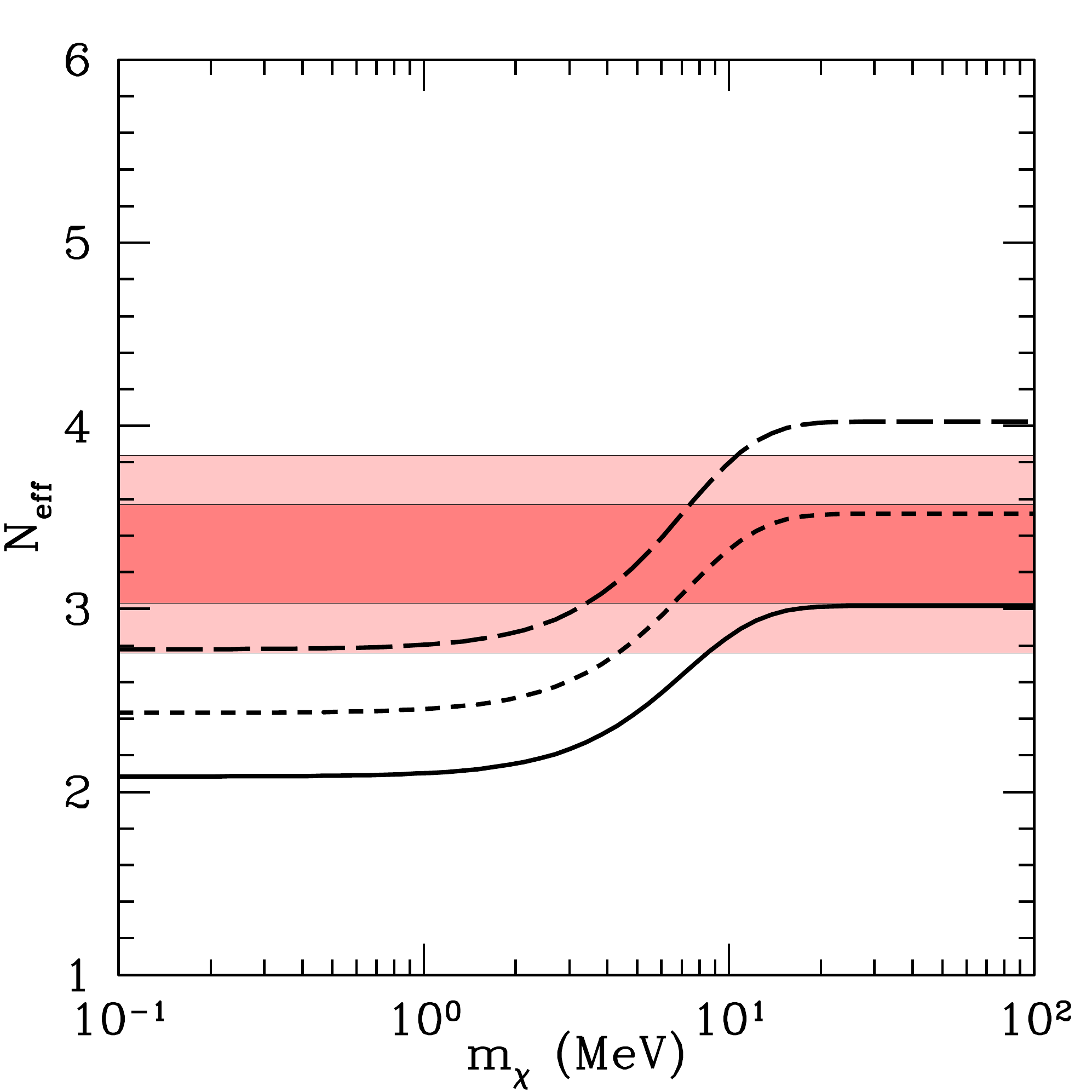}
\hskip .4in
\includegraphics[width=0.45\columnwidth]{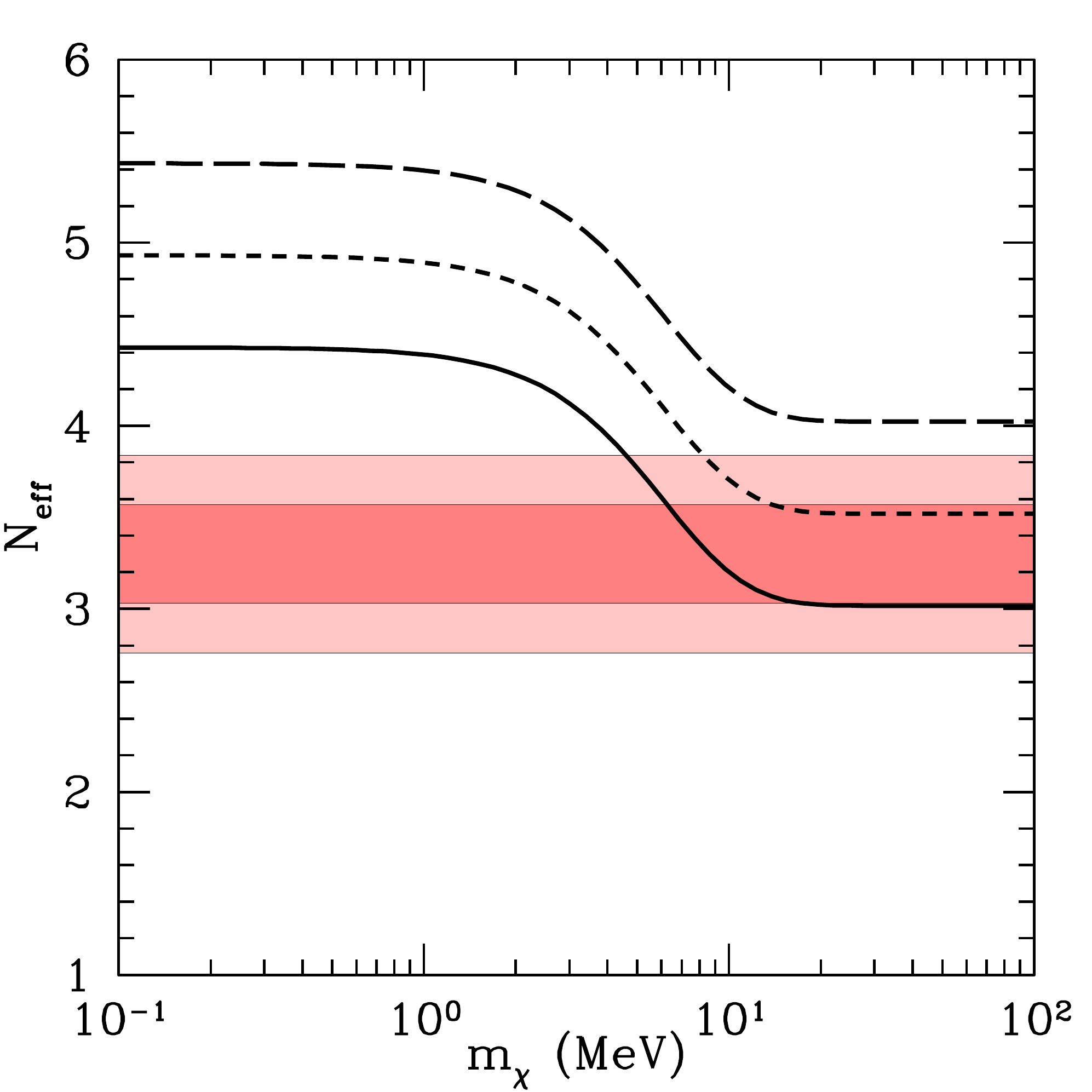}
\\\vskip 0.2in
\caption{(Color online)  The left panel shows N$_{\rm eff}$ as a function of the WIMP mass for an electromagnetically coupled, Majorana fermion WIMP and \Deln~equivalent neutrinos.  The solid curve is for $\Delta {\rm N}_{\nu} = 0$, the short-dashed curve is for $\Delta {\rm N}_{\nu} = 1/2$, and the long-dashed curve is for $\Delta {\rm N}_{\nu} = 1$.  The horizontal, red/pink bands are the Planck CMB 68.3\% and 95.5\% allowed ranges for N$_{\rm eff}$.  The right panel shows the corresponding results for a neutrino coupled, Majorana fermion WIMP.}
\label{fig:neffvsmall}
\end{center}
\end{figure}

An ``equivalent neutrino", $\xi$, is a very light ($m_{\xi} \ll m_{e}$) particle that is still ER when $T = T_{\gamma 0}$ (\ie, at BBN and at recombination, but not necessarily at present) and is not included in the SM.  Such particles could be populated by mixing with the SM neutrinos (\eg, \cite{dodelsonwidrow}).  In this case, each Majorana equivalent neutrino contributes $\Deln \leq 1$.  Alternatively, they may have once been in equilibrium with the SM particles at a high temperature, but decoupled before the SM neutrinos decoupled, at a temperature $T_{\xi d} \geq T_{\nu d}$.  In either case, it is assumed that the equivalent neutrinos are already decoupled from all SM particles when the SM neutrinos decouple.  For WIMPs that annihilate exclusively to \epm pairs and photons, neither the SM nor equivalent neutrinos are heated after $T_{\nu d}$, and Eq.~(\ref{eq:neff0-nowimp}) for $\neff^{0}$ becomes a function $\neff^{0}(m_{\chi})$, differing from $\neff^{0} = 3.02$.  The $\neff - m_{\chi}$ relation is shown for three values of \Deln~in the left hand panel of Fig.\,\ref{fig:neffvsmall}\,.  For this case, BBN and the CMB were used in our previous paper \cite{kngs1} to constrain \neff~and \Deln, where it was found that since $\neff^{0} < 3$ in the presence of an EM coupled light WIMP, at low \mchi, $\Deln > 0$ is allowed, and even required, in order to match the measured \neff.

\subsection{Neutrino Coupled Light WIMPs}
\label{sec:nureview}

If, instead, the SM neutrinos are heated by the annihilation of neutrino coupled WIMPs, the case explored here, there are two possibilities.  If the WIMP annihilation heats SM neutrinos, but not the already decoupled equivalent neutrinos, then the $\neff - \Deln$ relation \cite{chimera} is modified, to
\beq
\neff = 3.02\bigg[\bigg(1 + {4\tilde{g}_{\chi}\phi_{\chi d} \over 21}\bigg)^{4/3} + {\Deln \over 3}\bigg]\,,
\label{eq:neff-nnu-caseA}
\eeq
where $\tilde{g}_{\chi} = 1,\,(7/4,\,2,\,7/2)$ if the WIMP is a real scalar (Majorana fermion, complex scalar, Dirac fermion).  Note that this equation has a different form than Eq.~\ref{eq:original-neff}.  \neff~for \Deln~= 0 (\ie, $\neff^{0}$) is shown as a function of the WIMP mass for neutrino coupled WIMPs by the upper set of curves in Fig.\,\ref{fig:neff0vsmall}.  \neff~is shown as a function of the WIMP mass for a Majorana WIMP, for nonzero values of \Deln, in the right hand panel of Fig.\,\ref{fig:neffvsmall}.\footnote{The version of this graph that appeared in our previous paper \cite{kngs1} contained an error in the calculation of the $\Deln \neq 0$ curves.  This error affected nothing else in that paper.}

The other possibility is that the annihilating WIMP heats both the SM and the equivalent neutrinos.   In this case
\beq
\neff = 3.02\bigg(1 + {\Deln \over 3}\bigg)\bigg[1 + {4\tilde{g}_{\chi}\phi_{\chi d} \over 21 + 7 \Deln}\bigg]^{4/3}\,.
\label{eq:neff-nnu-caseC}
\eeq
Note that when \Deln~= 0, both results reduce to $\neff^{0} = 3.02(1 + 4\tilde{g}_{\chi}\phi_{\chi d}/21)^{4/3}$.  The difference between these two possibilities when $0 \leq \Deln \leq 1$ is so small that it is invisible on the scale of Fig.\,\ref{fig:neffvsmall}.  This reflects the physics involved: the entropy carried initially by the WIMPs is shared among more species in the second case, but each species receives less entropy.  As far as the expansion rate of the universe is concerned, these two effects very nearly cancel.   It will be seen below that since the two scenarios for WIMP coupling to neutrinos are equivalent if either $\mchi \rightarrow\infty$ (no light WIMP) or $\Deln = 0$ (no equivalent neutrinos), there is almost no difference between them in fitting the present data.  A curiosity of assuming equilibrium between the WIMPs and all neutrinos (SM plus equivalent) is that it implies a common temperature for all neutrino species and therefore restriction of \Deln~to integers or integer multiples of 4/7; in computing this scenario, we have kept \Deln~as a continuous parameter by allowing arbitrary real values for the spin degeneracy factor of the equivalent neutrinos.

As may be seen from Fig.\,\ref{fig:neffvsmall}, while $\Deln \gsim 1$ is allowed by the CMB for an EM coupled light WIMP, values of $\Deln \gsim 1$ are disfavored in the presence of a neutrino coupled light WIMP.  Indeed, even for \Deln~= 0, a neutrino coupled WIMP cannot be very light before coming into conflict with the CMB constraint on \neff.  This will become clearer when the BBN and CMB results are compared and combined.

\section{Updated Constraints With And Without An EM Coupled Light WIMP}
\label{sec:nowimp}

\begin{figure}[!t]
\begin{center}
\includegraphics[width=0.5\columnwidth]{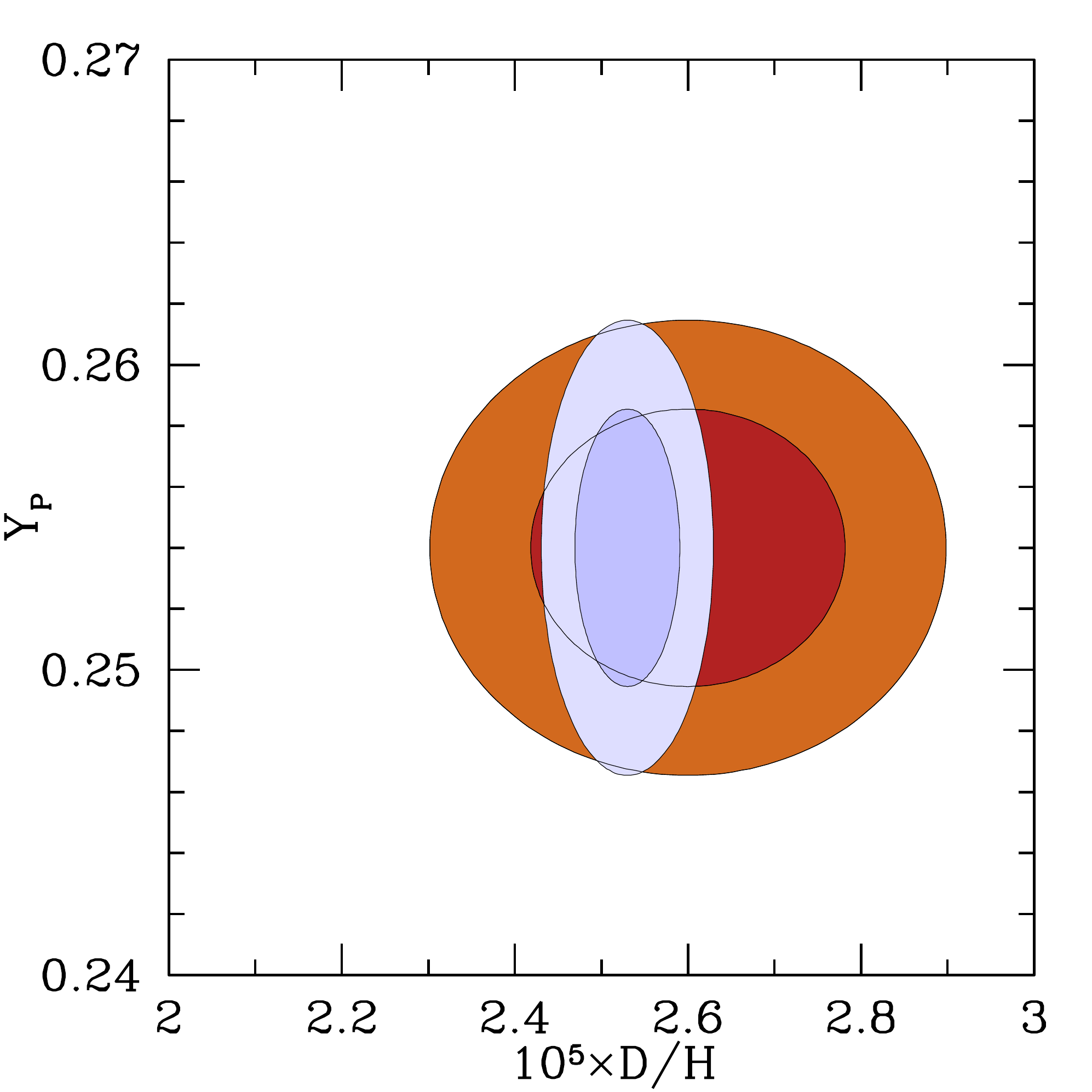}
\caption{(Color online)  The $68.3\,\%$ and $95.5\,\%$ likelihood contours of the observationally inferred primordial abundances of \4he and D in the \Yp~-- D/H plane.  The helium abundance, \Yp~$ = 0.254 \pm 0.003$, is adopted from Izotov \etal~\cite{izotov}.  The red/orange contours are for the older, Pettini \& Cooke \cite{pettini} D abundance, $y_{\rm DP} \equiv 10^{5}({\rm D/H})_{\rm P} = 2.60 \pm 0.12$, and the blue contours correspond to the newer, Cooke \etal~\cite{cooke} determination, $y_{\rm DP} = 2.53 \pm 0.04$.}
\label{fig:pvsc}
\end{center}
\end{figure}

In our previous paper \cite{kngs1} the consequences for BBN and the CMB, with and without a light WIMP that annihilates to \epm pairs and/or directly to photons, was explored.  Recently, there has been a new determination of the primordial deuterium abundance \cite{cooke}, whose central value is in excellent agreement with the value adopted in \cite{kngs1}, but whose uncertainty has been reduced by a factor of three.   Confidence level contours of the observationally determined primordial abundances of D and \4he are shown in Figure \ref{fig:pvsc}, comparing the older, Pettini \& Cooke D abundance \cite{pettini}, to the newer, Cooke \etal~value \cite{cooke}.  This small difference in the D abundance and the large reduction in its uncertainty lead to a small, but noticeable, effect on the central values, and a larger effect on the errors, in the observationally inferred model parameters.  In addition to the influence of the equivalent neutrino and WIMP energy densities, the BBN yields depend on the baryon density of the universe, expressible as the ratio, by number, of baryons to photons, $\eta \equiv 10^{10}\eta_{10}$.  This ratio and the baryon mass density parameter \omb~are related by $\eta_{10} \approx 273.9\,\omb$ \cite{steigman2006}.   To a good approximation, adopting the new D abundance determination in a BBN-only fit increases $\eta_{10}$ by $\sim 0.1$ (\omb~increases by $\sim 0.0004$) and results in a decrease in \Deln~(and \neff) of order $\sim 0.01$ \cite{cooke}.  In \S\,\ref{sec:update1} and \S\,\ref{sec:update2} the BBN and the joint BBN and CMB constraints are updated.

\subsection{Updated BBN And CMB Constraints In The Absence Of A WIMP}
\label{sec:update1}

\begin{figure}[!t]
\begin{center}
\includegraphics[width=0.32\columnwidth]{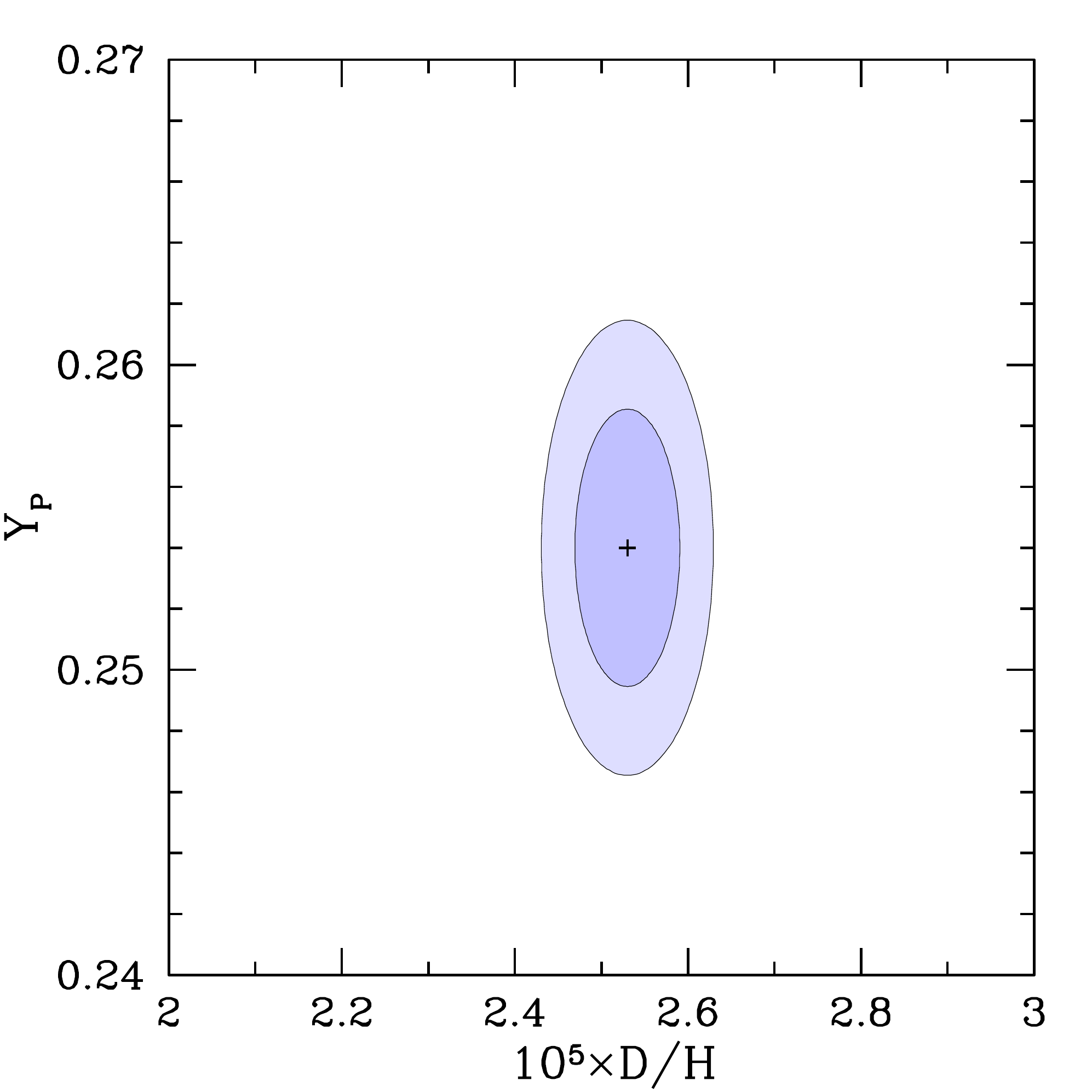}
\includegraphics[width=0.32\columnwidth]{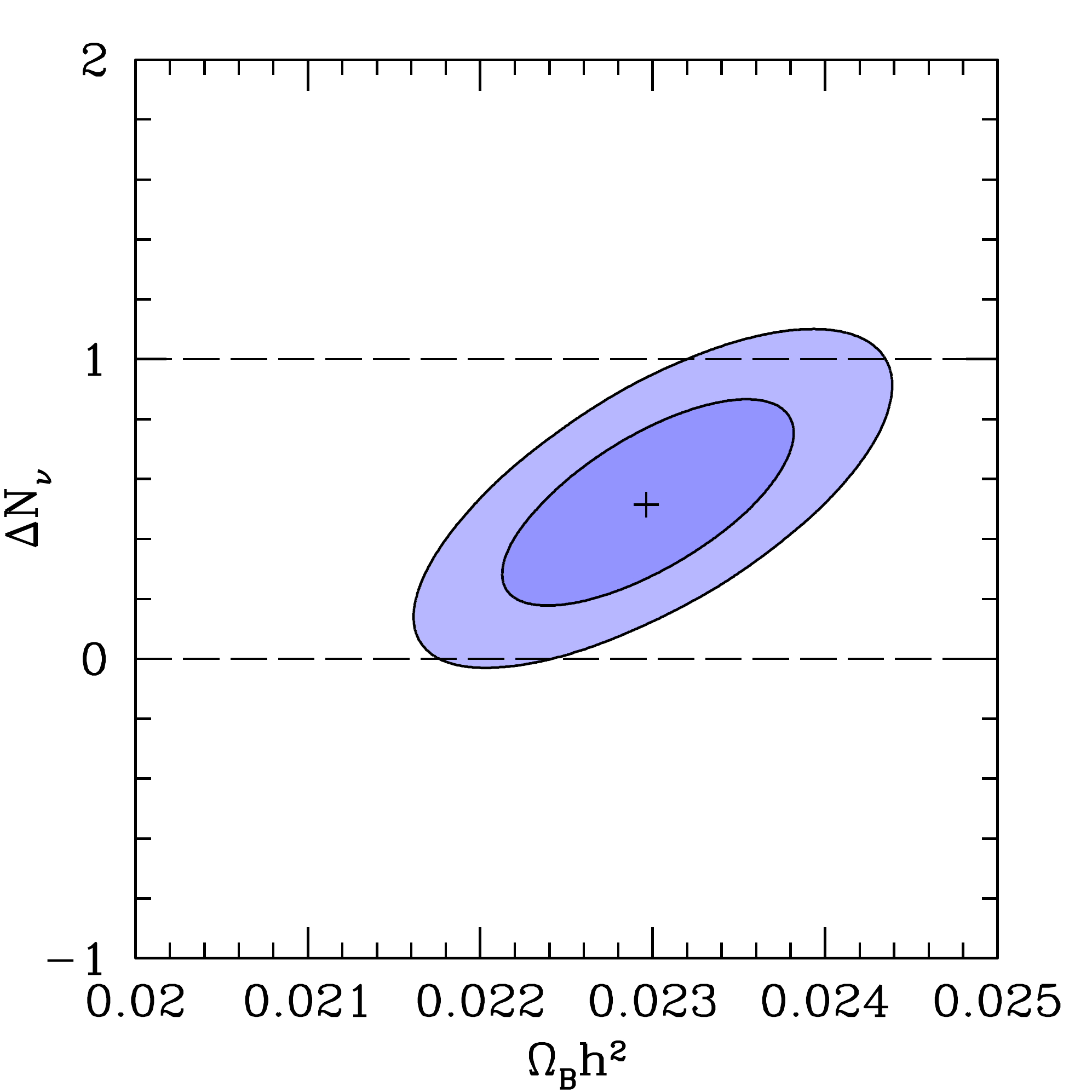}
\includegraphics[width=0.32\columnwidth]{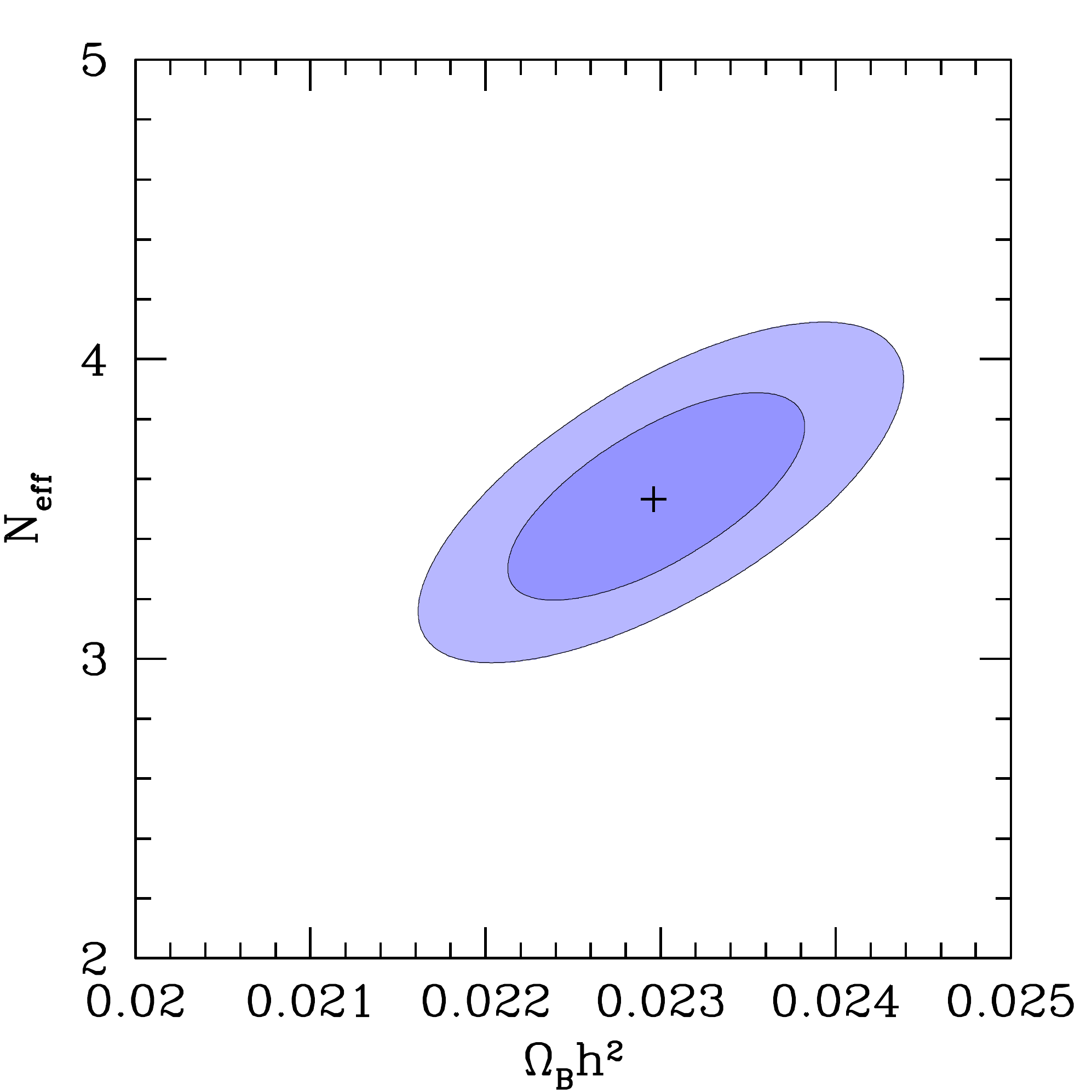}
\caption{(Color online) The left hand panel shows the 68.3\% and 95.5\% likelihood contours in the helium abundance -- deuterium abundance  (\Yp~-- $y_{\rm DP}$) plane.  The black cross corresponds to \Yp~= 0.254~\cite{izotov} and $10^{5}({\rm D/H})_{\rm P} = 2.53$~\cite{cooke}.  See the text for details.  The middle panel shows the corresponding BBN-inferred 68.3\% and 95.5\% contours for \Deln~and \omb, in the absence of a light WIMP and in the IND approximation.  The dashed horizontal lines show \Deln~= 0 and \Deln~= 1 as guides to the eye.  The right hand panel shows the corresponding contours for \neff~and \omb.  The black crosses in the middle and right hand panels show the best fit BBN values; without the IND approximation, they would fall at \Deln~= 0.50, \neff~= 3.56, and \omb~= 0.0229.}
\label{fig:bbnonly}
\end{center}
\end{figure}

First, assume that there is no light WIMP.  Recall that for this case, $\neff = 3.05(1 + \Deln/3)$.  In the left hand panel of Figure \ref{fig:bbnonly} the new limits are shown in the \Yp~-- D/H plane.  For BBN only (no CMB constraints), in the absence of a light WIMP (EM or neutrino coupled), this pair of observables, \{\Yp,\,{\rm D/H}\}, may be mapped exactly into the parameter pair, \{\neff,\,\omb\}, as shown in the right hand panel of Fig.\,\ref{fig:bbnonly}\,\footnote{Because the BBN code used here employs the IND approximation, all figures show results consistent with using $\neff^{0} = 3.02$ as the numerical constant in Eqs.~(\ref{eq:neff-nnu-caseA}) and (\ref{eq:neff-nnu-caseC}).  When there is no light WIMP, it is a reasonable approximation to replace 3.02 by 3.05 and to include the related effect on \Yp~as computed in Ref.~\cite{mangano} with $\Deln = 0$.  Accordingly, for the no-WIMP cases, we quote numerical results using $\neff^0 = 3.05$ instead of 3.02 and have applied a small correction to the computed \Yp.}.  For BBN only, using the new D abundance, 
\begin{eqnarray}
\Deln & = & 0.50 \pm 0.23\\\nonumber
\neff & = & 3.56 \pm 0.23 \\\nonumber
\eta_{10} & = & 6.28 \pm 0.15~~~
(100\,\omb  =  2.29 \pm 0.06)\,.
\end{eqnarray}
The current D and \4he abundances favor neither standard big bang nucleosynthesis (SBBN: \Deln~= 0), nor the presence of a  thermally equilibrated sterile neutrino (\Deln~= 1).

\begin{figure}[!t]
\begin{center}
\includegraphics[width=0.32\columnwidth]{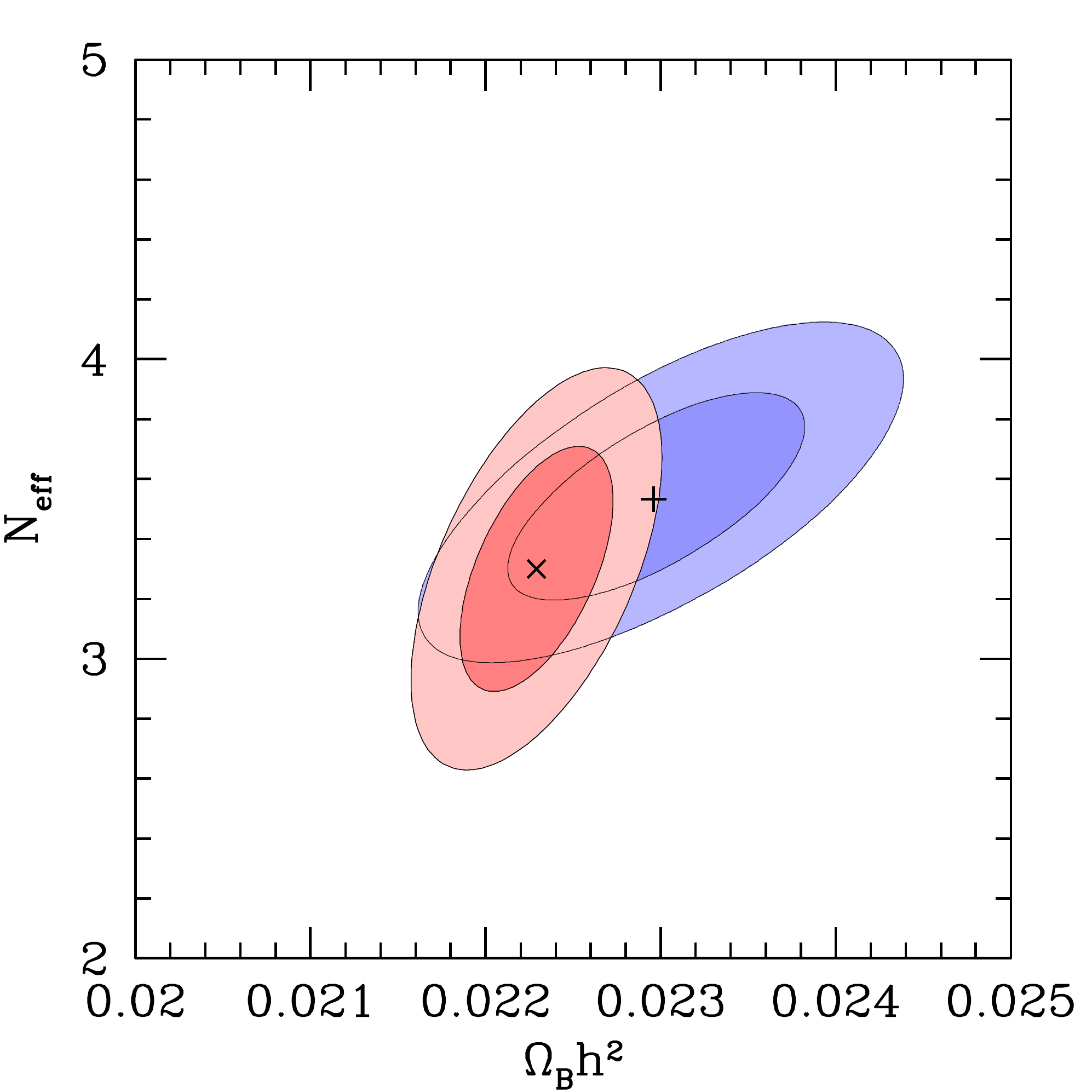}
\hskip .4in
\includegraphics[width=0.32\columnwidth]{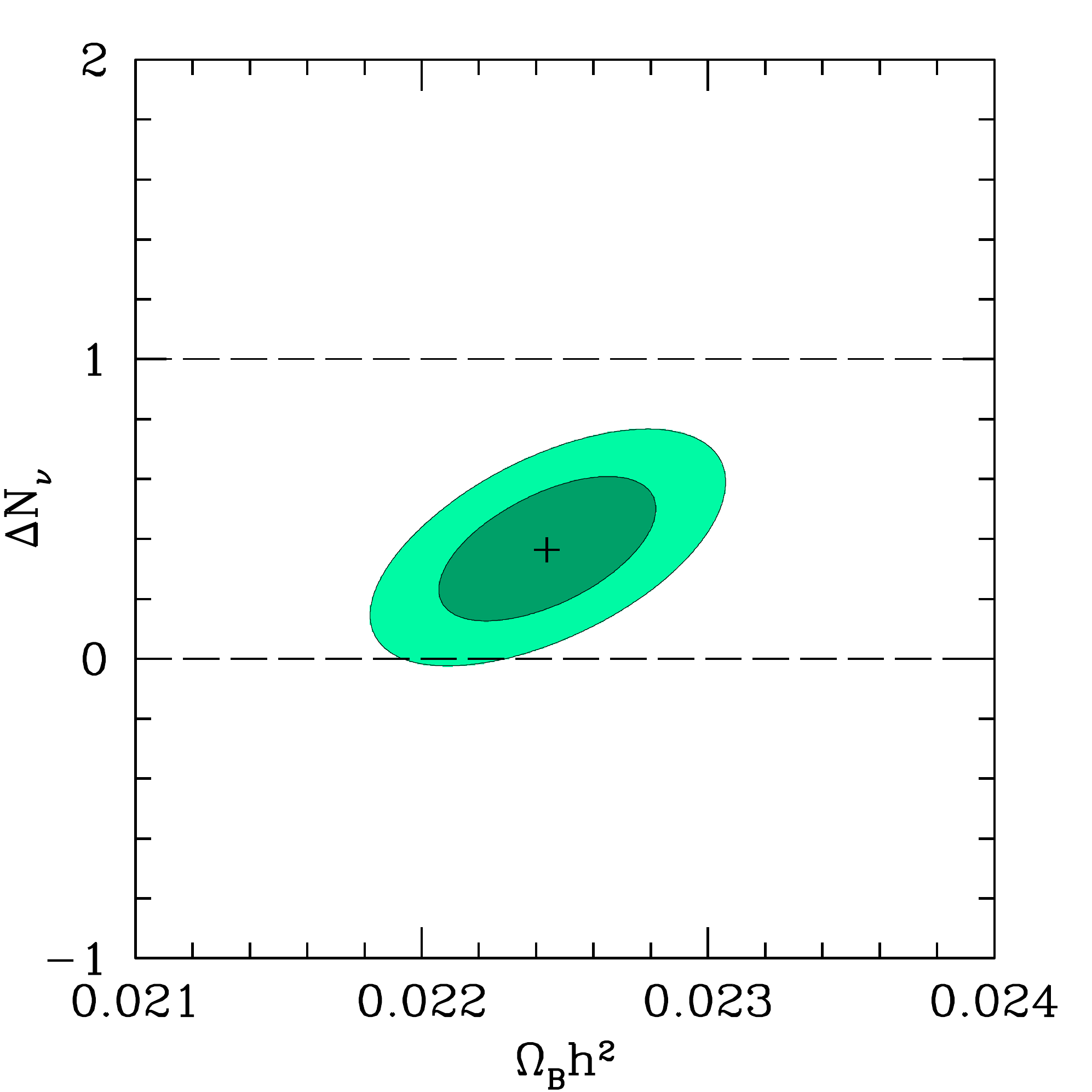}
\\\vskip 0.2in
\caption{(Color online) The left hand panel compares the BBN (upper right, blue contours) and CMB (lower left, pink contours) constraints in the \neff~-- \omb~plane.  The ``x" shows the best fit CMB point and the ``+" shows the best fit BBN point.  In the right hand panel the joint BBN + CMB fit is shown in the \Deln~-- \omb~plane.    The ``+" shows the best joint fit point, which is located at \Deln~= 0.35 and \omb~= 0.0224 (without resorting to the IND approximation).}
\label{fig:bbncmb}
\end{center}
\end{figure}

In the left hand panel of Figure \ref{fig:bbncmb} the independent BBN and CMB likelihood contours in the $\neff - \omb$ plane are compared.  The Planck results (those described by Eq.~74 of Ref.~\cite{planck}) are adopted and are represented as shown in Table II of Ref.~\cite{kngs1}.   Despite the BBN contours having been displaced by the new D/H value to slightly higher \omb, there is still very good agreement between BBN and the CMB ($\chi^2 = 1.11$ with 2 degrees of freedom), justifying a joint BBN + CMB fit.  The results of that joint fit are shown in the right hand panel of Fig.\,\ref{fig:bbncmb}, where the 68.3\% and 95.5\% contours are shown in the $\Deln - \omb$ plane.  
The new, joint BBN + CMB fit, parameter values are
\begin{eqnarray}
\label{eq:bbn+cmb-nowimp-fit}
\Deln & = & 0.35 \pm 0.16\\\nonumber
\neff & = & 3.40 \pm 0.16\\\nonumber
\eta_{10} & = & 6.15 \pm 0.07~~~~(100\,\omb = 2.24 \pm 0.03)\,.
\end{eqnarray}
Thus, while the new D abundance still allows \Deln~= 0 at $2.2\,\sigma$ ($\Deln \geq 0$ is favored at $\gsim 98.8\%$ confidence), it strongly disfavors even one sterile neutrino, at $\sim 4.0\,\sigma$ ($\Deln > 1$ excluded at $> 99.9\%$ confidence).

\subsection{Constraints On New Physics Between BBN And Recombination}
\label{newphys}

Since BBN is sensitive to physics when the Universe was a few minutes old, and the CMB probes physics some 400 thousand years later, the excellent agreement between the BBN and CMB constraints on the cosmological parameters explored here constrains at least some possibilities for ``new physics" between these epochs in the early evolution of the Universe.  If it is assumed that the ``physics beyond the SM" that may arise between BBN and recombination can be described by the differences between the baryon to photon ratio ($\eta_{\rm B} = 10^{-10}(n_{\rm B}/n_{\gamma}$) and \Deln~evaluated from BBN and the CMB, then their current agreement, within the errors, limits some classes of nonstandard physics.  For example, if the baryon number is conserved in this interval, then the ratio $\eta_{10}^{\rm BBN}/\eta_{10}^{\rm CMB} = N_{\gamma}^{\rm CMB}/N_{\gamma}^{\rm BBN}$ probes the difference between the number of photons in a comoving volume at recombination ($N_{\gamma}^{\rm CMB}$) compared to the number present at the end of BBN ($N_{\gamma}^{\rm BBN}$).  While entropy conservation suggests that $N_{\gamma}^{\rm CMB}/N_{\gamma}^{\rm BBN} \geq 1$, any deviation of this ratio from unity measures (limits) photon production (\eg, from the decay or annihilation of a beyond the standard model particle) between these two, widely separated epochs.  Since the standard models of particle physics and cosmology lack equivalent neutrinos, \Deln~probes any ``extra" (beyond the SM) contribution to the early Universe radiation energy density (``dark radiation") at BBN and at recombination (or, at the epoch of equal matter and radiation densities).  A nonzero value of the change in \Deln, $\delta(\Deln) = \Deln^{\rm BBN} - \Deln^{\rm CMB}$, could also signal exotic, new physics.  It is evident from Fig.~\ref{fig:bbncmb} that both of these difference parameters, $\Delta N_{\gamma}/N_{\gamma}^{\rm BBN} = N_{\gamma}^{\rm CMB}/N_{\gamma}^{\rm BBN} - 1$ and $\delta(\Deln) = \Deln^\mathrm{BBN} - \Deln^\mathrm{CMB}$, are zero within the uncertainties.  Fitting these two parameters results in $\Delta N_{\gamma}/N_{\gamma}^{\rm BBN} = 0.027 \pm 0.026$ and $\delta(\Deln) = 0.25 \pm 0.35$.  Since we know that $\Delta N_{\gamma} \geq 0$, it probably makes more sense to compute an upper limit for $\Delta N_{\gamma}/N_{\gamma}^{\rm BBN}$.  The one-sided 95.5\% limit (so that 4.5\% of the likelihood lies at higher values) is $\Delta N_{\gamma}/N_{\gamma}^{\rm BBN} < 0.072$.  Notice that since the CMB displays a thermal spectrum, any photons produced after BBN must have been thermalized well before recombination.  This has been assumed here.  If thermalization occurred at redshift $5\times 10^4 < z < 2\times 10^6$ \cite{chluba}, this would have occurred at fixed photon number, and the CMB frequency spectrum would show a $\mu$-distortion (nonzero chemical potential).  There is a tight observational limit of $|\mu| < 9\times 10^{-5}$ \cite{cobe-mu}, strictly limiting the addition of photons at $z < 2\times 10^6$.

\subsection{Updated BBN And CMB Constraints On An EM Coupled Light WIMP}
\label{sec:update2}

The new D abundance has a nearly negligible effect on the combined BBN and CMB parameter constraints for a light WIMP that annihilates to \epm pairs and/or photons in chemical equilibrium.  The parameter constraints remain very nearly independent of the statistics obeyed by the WIMP (fermion or boson), with the largest variation among WIMP types occurring in the lower limits on the WIMP mass, from $\sim 0.5$ MeV for a real scalar WIMP to $\sim 5$ MeV for a Dirac fermion WIMP\,\footnote{As noted in \cite{kngs1}, the variation in the best fit mass values is much smaller, ranging from $\sim 5$ MeV for a real scalar to $\sim 10$ MeV for a Dirac fermion.}.  In light of the new data, the presence of a light EM coupled WIMP is still a good fit ($\chi^2 \sim 0.5$ for 1 degree of freedom) and remains slightly favored ($\Delta \chi^2 \sim 0.7$) over its absence; the best fit WIMP mass decreases from its previous value by $\lsim 1\,{\rm MeV}$ (\eg, for a Majorana fermion WIMP, $\Delta m_{\chi} \approx -\,0.8\,{\rm MeV}$), while the $\mchi \rightarrow \infty$ limit remains a good fit.   For the other parameters, \neff~decreases slightly, from $\neff = 3.30 \pm 0.26$ to $\neff = 3.22\pm 0.25$, and \Deln~is virtually unchanged, $\Deln = 0.65^{+0.45}_{-0.37}$.  The lower limits on the WIMP masses are loosened by only $\sim 0.2$ MeV for Majorana and $\sim 0.3$ MeV for Dirac WIMPs \footnote{We note that the previous paper suffered from an error that placed the lower limits on \mchi~at 96\% confidence rather than 95.5\%, so that the tabulated limits were 0.1 to 0.2 MeV lower than intended.  The further correction due to the revised D/H is independent of this and in the opposite direction, so that the revised D/H mainly cancels the mistake.}.

\section{BBN In The Presence Of A Neutrino Coupled Light WIMP}
\label{sec:bbnnuwimp}

\subsection{Physical effects}

For BBN, the case considered here, of an equilibrated light WIMP that couples only to neutrinos, is in some ways the opposite of the case examined in Ref.~\cite{kngs1}, where the light WIMP coupled to the electromagnetic plasma.  In both cases, the light WIMP speeds up expansion at very early times by contributing directly to the energy density.  As the temperature drops and the WIMP becomes nonrelativistic, the entropy of a neutrino-coupled WIMP is transferred to the neutrinos, giving them a higher temperature at fixed $T_\gamma$ than in the standard model.  This ensures that the expansion rate (again, at fixed $T_\gamma$) is faster than in the standard model.  In addition, in this case the neutrino heating skews the weak interactions interconverting protons and neutrons, because hotter neutrinos more easily supply the $Q$-value to make a neutron and a positron from a proton.  These effects, linked to neutrino temperature, are the opposite of the electromagnetically coupled case.

\begin{figure}[!t]
\begin{center}
\includegraphics[width=0.5\columnwidth]{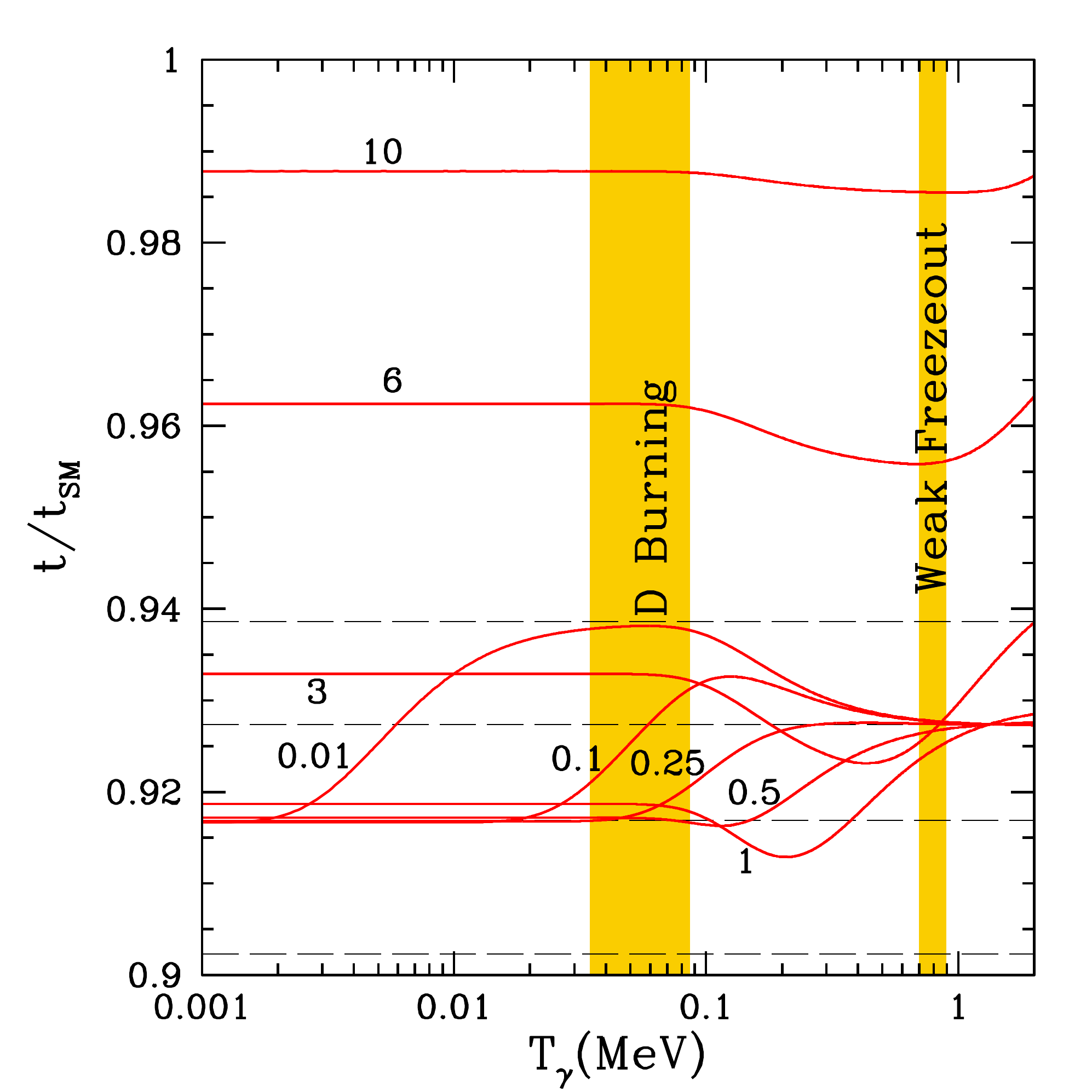}
\caption{(Color online) The expansion timescale (reciprocal of the logarithmic time derivative of the scale factor) is shown as a function of the photon temperature ($T_\gamma$) for several values of the mass (\mchi) of a light Majorana WIMP.  Timescales are shown in units of $t_\mathrm{SM}$, the timescale at the same  $T_\gamma$ in the standard model.  Curves are labeled by \mchi~in MeV.  Vertical bands indicate the temperatures at which weak freezeout of the nucleons and final burning of deuterium (and production of lithium) occur.  From top to bottom, horizontal dashed lines indicate limits in which \epm have annihilated but WIMPs have not; \epm and WIMPs are both fully relativistic; \epm and WIMPs have fully annihilated; and WIMPs have annihilated fully but the \epm pairs have not.  The bottom dashed line is not reached because $m_e$ and $T_{\nu d}$ are so close that it is not possible for WIMPs of any mass to go from the relativistic limit at neutrino decoupling to fully annihilated before \epm annihilation.}
\label{fig:timescales}
\end{center}
\end{figure}

The effects on the expansion rate are shown in Fig.~\ref{fig:timescales}, in which the expansion timescale (comparable to the age of the universe) is shown as a function of $T_\gamma$ for Majorana WIMPs of several masses, assuming that neutrinos decouple suddenly from the EM plasma at $T_{\nu d} =2$ MeV.  Since the timescale is shown normalized to its value, $t_\mathrm{SM}$, at the same $T_\gamma$ in the standard model, values below unity reflect expansion that has been sped up by the light WIMP.  A 10 MeV WIMP (the highest mass shown) contributes some energy density of its own to the expansion rate at early times but quickly transfers that energy to the neutrinos.  As the WIMP mass decreases, more of its entropy ends up being carried by neutrinos, but the transfer happens later.  WIMPs with $\mchi \lesssim T_{\nu d}$ contribute initially as a relativistic species, equivalent to a unit increment of \Deln~if the WIMP is Majorana; this limit is shown as the second-highest dashed line in Fig.~\ref{fig:timescales}.  Once the WIMPs and the electrons and positrons have all annihilated, a Majorana WIMP has heated the neutrinos so that the expansion rate is about 8\% faster than in the standard model -- this limit is the second-lowest dashed line in Fig.~\ref{fig:timescales}.  If $\mchi \lesssim m_e$, there is an intermediate time at $T_\gamma \lesssim m_e$ when electrons and positrons have annihilated but WIMPs have not.  In Fig.~\ref{fig:timescales} this is visible as a rise in timescale (slower expansion) toward a limit that remains about 6\% faster than standard-model expansion.  The temperature range covered by this rise is exactly the range where the deuterium and lithium abundances are determined, late in BBN; it will be shown below that this results in a reversal of the general trend of abundances as functions of \mchi.

One effect found in the electromagnetically coupled case has no analogue here: the annihilation of an EM coupled WIMP ultimately produces photons, changing the baryon to photon ratio $\eta$.  WIMP annihilation to photons after \epm annihilation (especially if it happens after BBN ends) produces large shifts in the BBN yields at fixed late-time $\eta$, and this dominates much of the BBN results in Ref.~\cite{kngs1}.  The effects on BBN of a neutrino-coupled light WIMP therefore tend to be generally smaller than those of an EM coupled light WIMP.

\subsection{BBN calculations}
\label{sec:bbn-calc}

The effects described above have been included in BBN calculations using a modified version of the Kawano BBN code \cite{kawano88,kawano92}, assuming IND at $T_{\nu d} = 2$ MeV.  With the exception of adding neutrino coupled WIMPs (and setting the $g$-factor of EM coupled WIMPs to zero), the code is as described in Ref.~\cite{kngs1}.  The interested reader is referred there for computational details.  Two details that are particularly important for the present results (though unchanged from the previous calculation) are mentioned here, before discussion of the new modifications.

The present results are based on a neutron lifetime of $\tau_n = 880.1\pm 1.1$ s (with this error propagated through the analysis), as recommended by the Particle Data Group in 2012
\cite{pdg2012}, though the recommended lifetime has now been revised to $880.3\pm 1.1$ s after reanalysis of three experiments \cite{pdg2014}.  Since this shift is smaller than the error estimate, and changing $\tau_n$ would produce confusing inconsistencies with our earlier work, we use the older value here.  Both  $\tau_n$ values are averages of discrepant data (with inflated error bars to reflect this); we hope that the disagreement will be resolved by the next generation of experimental results.  One indication of how far the results might shift is given by the current difference between the two experimental approaches, with the mean $\tau_n$ from  ``bottle'' methods being near the value used here and that from ``beam'' methods being larger by about 8 seconds \cite{wietfeldt-greene}.  This 8 second difference corresponds to about the same effect on BBN predictions  of \Yp~as a change of $-0.13$ in \Deln, so the main effect of even a drastically longer lifetime (for fixed \Yp) would likely be to reduce the fitted \neff~and \Deln~values by about this amount.  This shift is about half the size of the quoted $1\,\sigma$ uncertainties of these parameters, \eg~in \S\,\ref{sec:nowimp}.

Another important detail is the rate for the deuterium-destroying reaction $d(p,\gamma)^3\mathrm{He}$.  Most calculations incorporate an empirically-derived cross section \cite{ma97,solarfusion} for this process, which depends on a single experiment at the most important energies for BBN.  The calculations here are based on an {\it ab initio} calculation of the cross section \cite{viviani00,marcucci05}, which is likely to be more reliable (with a 7\% error assigned and propagated through our analysis) \cite{nollett11}.  The difference amounts to about a 5\% lower predicted D/H at fixed parameters here compared to much of the recent literature, and thus a roughly 3\% shift in the BBN inferred baryon mass density parameter, \omb.  Since there is good agreement between D/H and the CMB in Sec.~\ref{sec:update1}, the empirical rate implies tension between BBN and the CMB at about the 90\% confidence level when the empirical rate is used, as found in Refs.~\cite{divalentino14,salvati14}.  Use of the empirical rate shifts \Deln~downward by about 0.02 in the absence of CMB constraints.  In joint BBN and CMB fitting, \Deln~and \neff~shift downward by about 0.1, allowing $\Deln = 0$ at 95\% confidence, while the lower bounds on \mchi~tighten by about 10\%.  Further experimental  and theoretical work on this cross section in the 50 -- 500 keV  energy range is highly desirable.

Incorporation of a neutrino-coupled WIMP into the code required first promoting $T_\nu$ to a dynamic variable.  In the original Kawano code, the neutrino temperature used in the weak rates is computed from the baryon density, using the proportionality of both quantities to the cube of the expansion scale factor.  This works because the Kawano code assumes IND and starts after neutrino decoupling, so the neutrino energies simply redshift with the expansion of the Universe.

With the addition of neutrino-coupled WIMPs, $T_\nu$ satisfies a differential equation that must be solved simultaneously with that for the expansion, just like $T_\gamma$.  Where the $T_\gamma$ equation has terms for photon and \epm entropies, the $T_\nu$ equation has terms for neutrino and light WIMP entropies.  Since the WIMPs are assumed to be in equilibrium with the SM neutrinos, they share the same temperature.  The energy density of equivalent neutrinos is computed from its scaling with the baryon density just as before.  If the light WIMPs are in equilibrium with all neutrino species (as in Eq.~(\ref{eq:neff-nnu-caseC})), this can be accounted for by adjusting the spin-degeneracy factor of the ``SM neutrino'' term in the $T_\nu$ equation so that the total energy density of neutrinos is $(3+\Deln)\rho_\nu$, and setting the equivalent-neutrino density variable in the code to zero.  As mentioned above, we assumed even in this latter case that \Deln~can take on any value, so that the corresponding neutrino spin-degeneracy factor need not be an integer.

\begin{figure}[!t]
\includegraphics[width=0.49\columnwidth,angle=0]{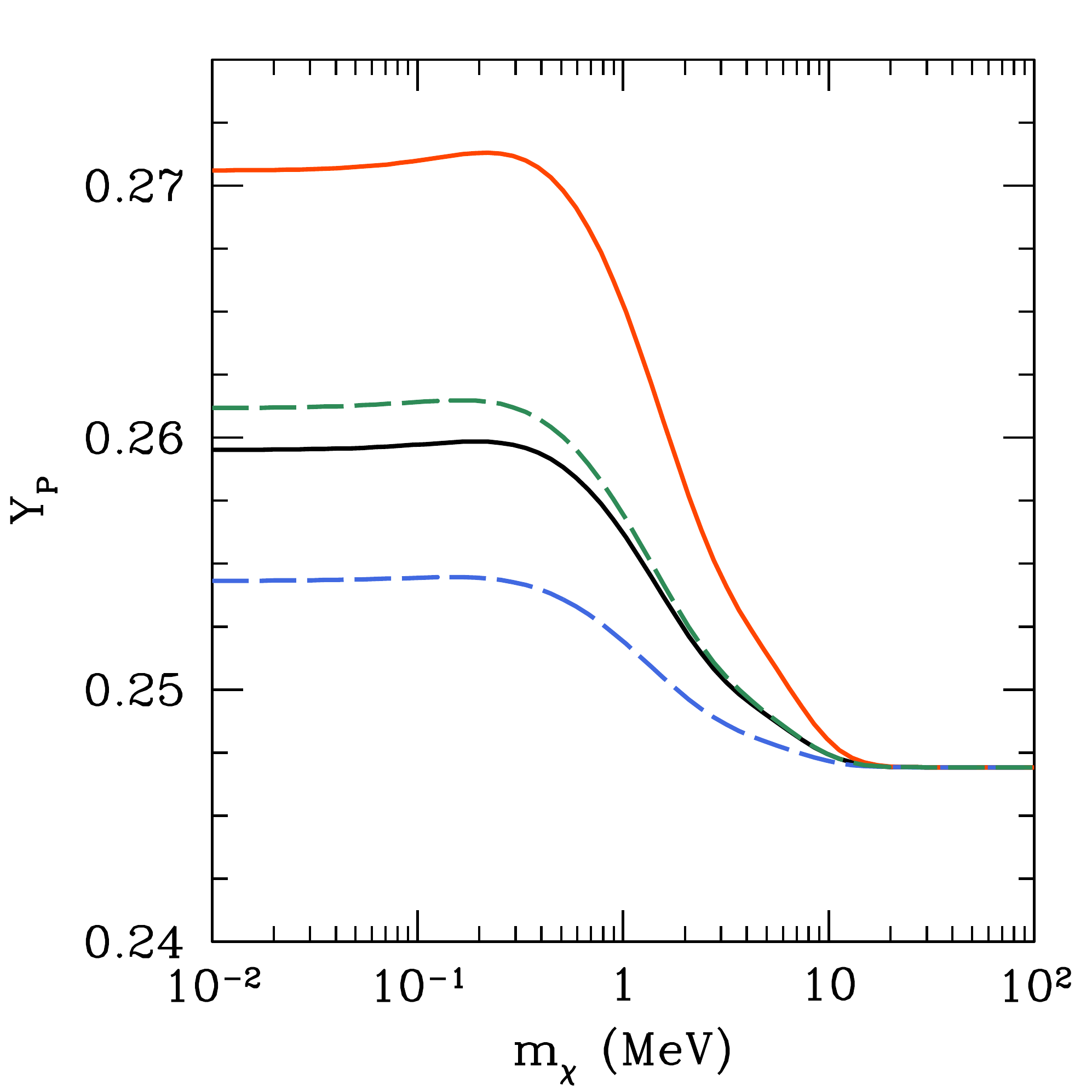}
\includegraphics[width=0.49\columnwidth,angle=0]{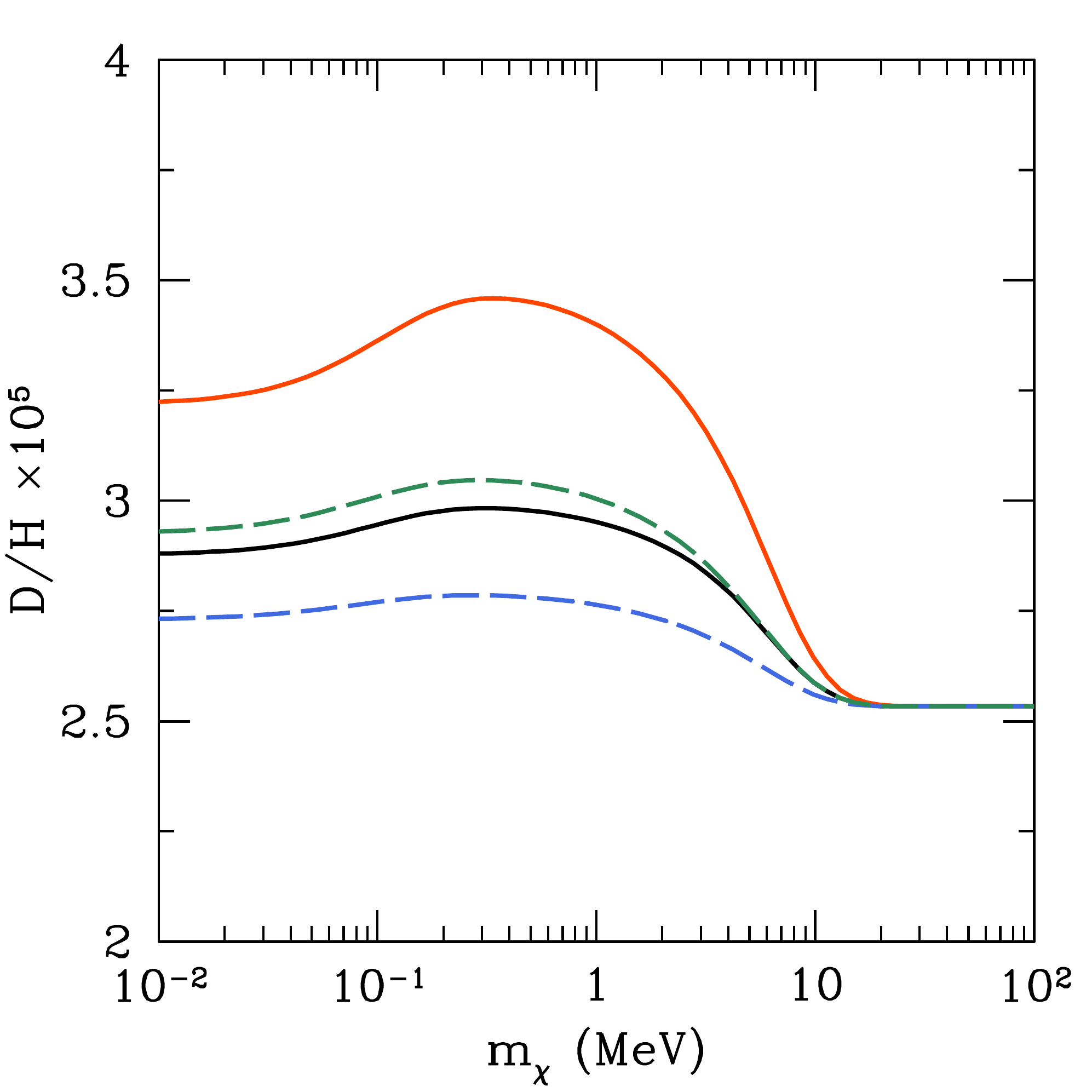}\\
\includegraphics[width=0.49\columnwidth,angle=0]{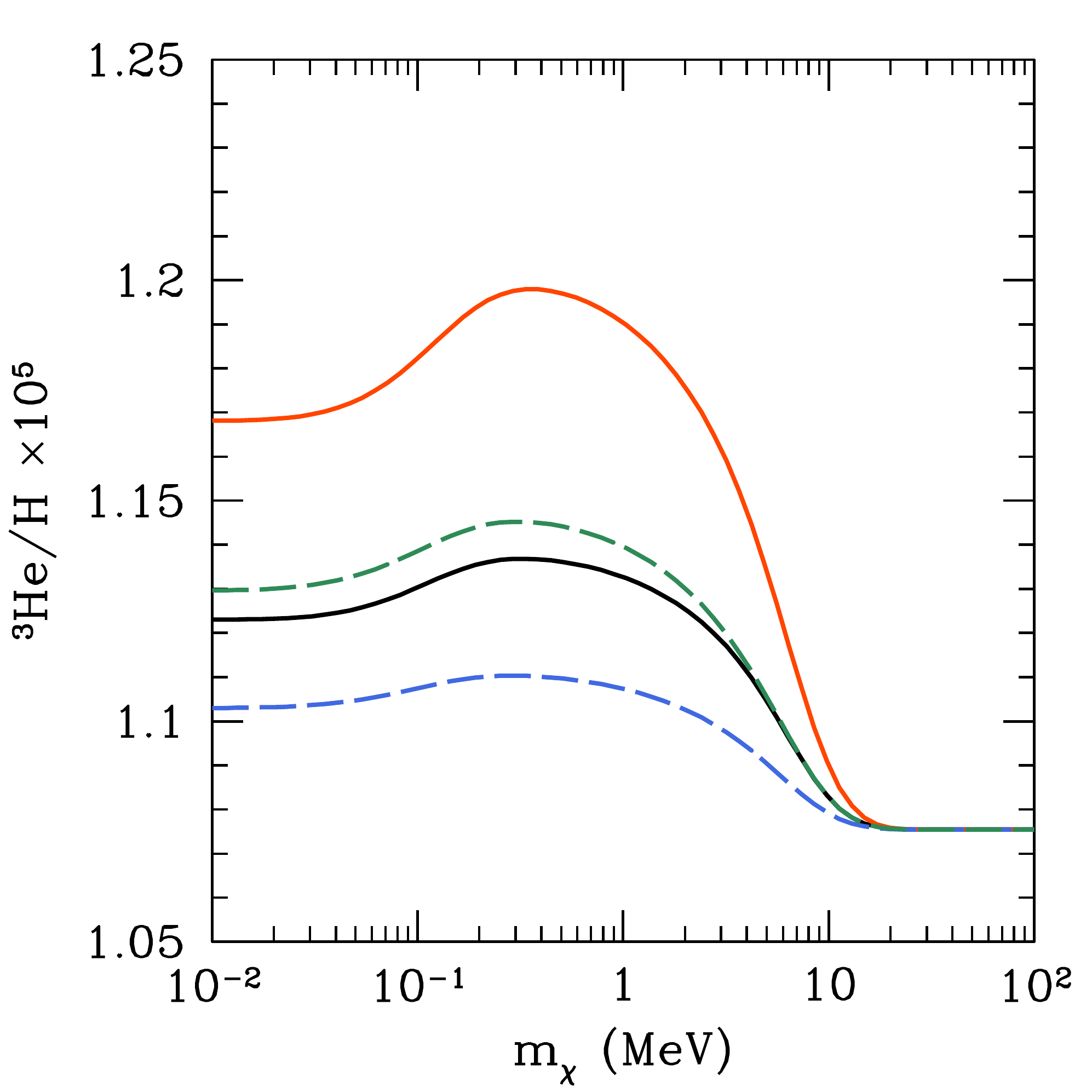}
\includegraphics[width=0.49\columnwidth,angle=0]{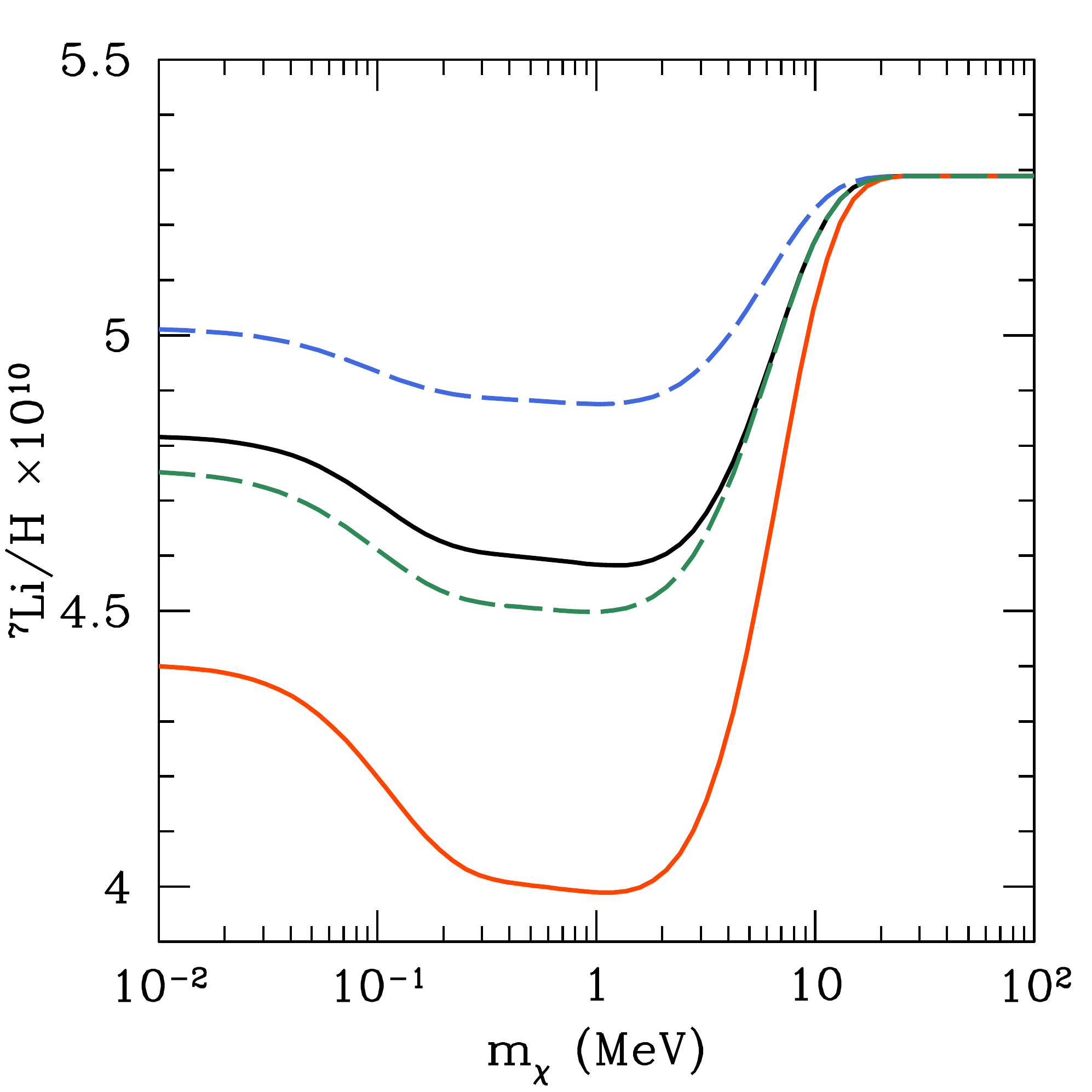}
\caption{(Color online)  For the case of a light WIMP that annihilates to the SM neutrinos, the four panels show the BBN yields of $^4$He (upper left), D (upper right), $^3$He (lower left), and $^7$Li (lower right) as functions of the WIMP mass, \mchi, for $\Omega_Bh^2 = 0.0220$ and $\Deln = 0$.  Solid curves show results for fermionic WIMPs (red for Dirac, black for Majorana) and dashed curves for bosonic WIMPs (green for a complex scalar, blue for a real scalar).  The sequence of curves is Dirac, complex scalar, Majorana, and real scalar, from top down in all panels except the lower right, where it is reversed.  The $^4$He abundance is shown as a mass fraction Y$_{\rm P}$, and the other abundances are shown as ratios by number to hydrogen.}
\label{fig:Deln=0yields}
\end{figure}

\begin{figure}
\includegraphics[width=0.5\columnwidth]{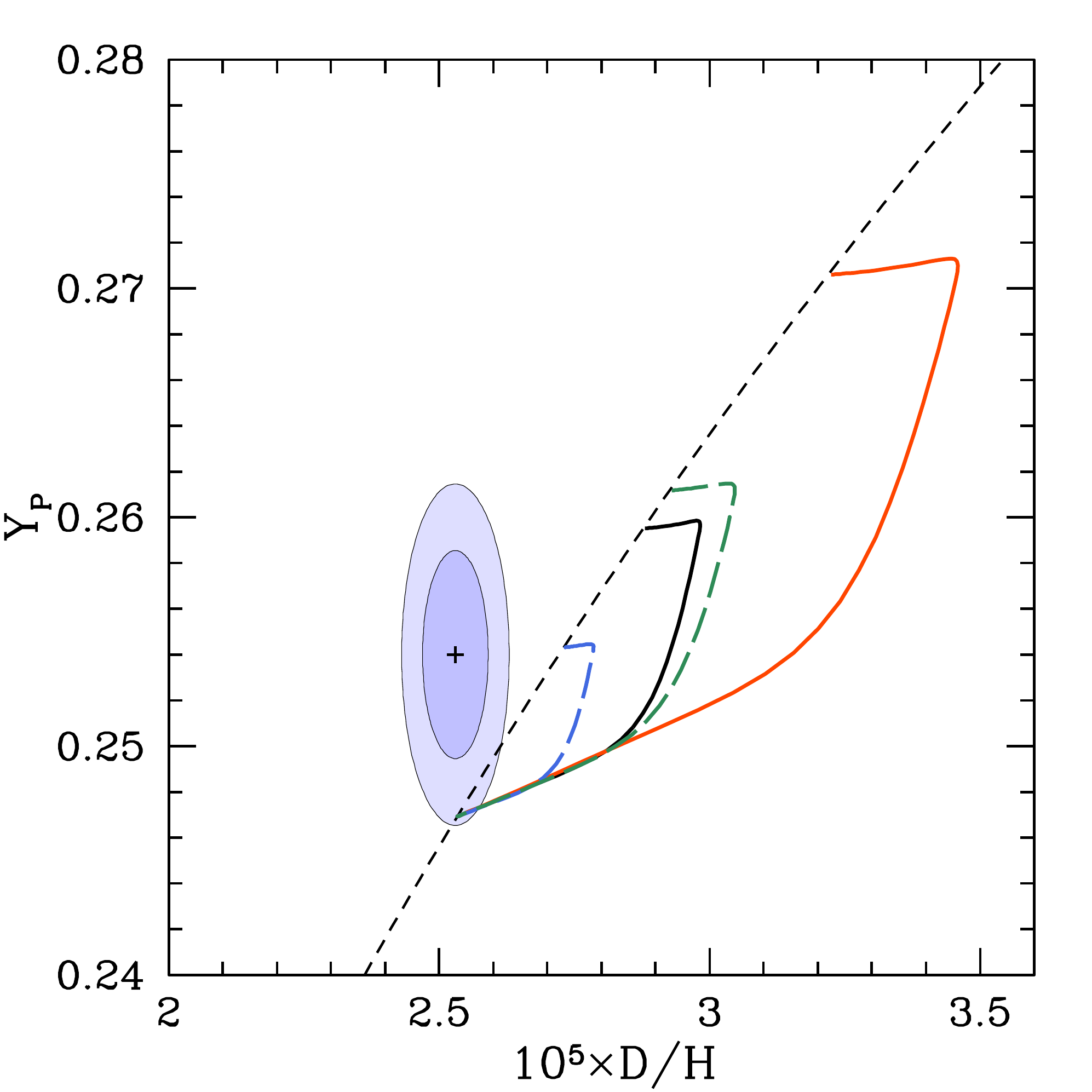}
  \caption{(Color online)  Yield curves for the \4he mass fraction, \Yp, and the deuterium abundance, $y_{\rm DP} = 10^{5}({\rm D/H})_{\rm DP}$,  for \Deln~= 0 and fixed \omb~= 0.0220, for the four types of of neutrino coupled WIMPs considered here: real and complex scalars (long-dashed curves, from left to right) and Majorana and Dirac fermions (solid curves, from left to right); colors as in Fig.\,\ref{fig:Deln=0yields}.  The WIMP mass decreases along the curves from the lower left (high mass, the no light WIMP limit) to the upper right (very light WIMP).  The short-dashed curve shows the no-WIMP limit at varying \Deln; the other curves each touch this limiting curve at both ends because because a WIMP with $\mchi \ll m_e$ is equivalent (for BBN) to a heavy WIMP (no light WIMP) with an incremented value of \Deln.   The 68.3\% and 95.5\% contours  for the D and \4he primordial abundances adopted here are also shown.}
  \label{fig:wimpyields-allkinds}
\end{figure}

The results of the BBN calculations including neutrino-coupled WIMPs are shown for all four WIMP spin-statistics possibilities considered and for $\Deln = 0$ in Figs.~\ref{fig:Deln=0yields} and \ref{fig:wimpyields-allkinds}.  The overall scale of a light WIMP's effects is smaller than in the corresponding figures of Ref.~\cite{kngs1}, where we considered an EM coupled WIMP.  These results are in good agreement with previously published calculations~\cite{ktw,serpico,boehm2012,boehm2013}.  The programming error suggested \cite{kngs1} as a possible explanation of the difference between Refs.~\cite{kngs1} and \cite{serpico} for EM coupled WIMPs, does not affect the calculations for neutrino-coupled WIMPs.

The results for D, $^3$He, and $^7$Li in Fig.~\ref{fig:Deln=0yields} can be understood entirely from the  timescales shown in Fig.~\ref{fig:timescales}.  For $\mchi \gtrsim 30$ MeV, the WIMP annihilates before neutrino decoupling and has no effect on the ratio of neutrino and photon temperatures.  For lower \mchi, the WIMP energy density initially contributes to the expansion rate and then raises $T_\nu/T_\gamma$ relative to the SM.   This causes expansion at fixed $T_\gamma$ to be faster, including at $T_\gamma \sim 50$ keV when the D, $^3$He, and $^7$Li abundances are determined.  Since D and $^3$He are being burned at the end of BBN, faster expansion implies less time to burn them away, leading to higher relic abundances.  The $^7$Be precursor of $^7$Li is still being produced at the end of BBN, so faster expansion implies less time for its production and therefore results in a lower relic $^7$Li abundance.

If the WIMPs are so light that they do not annihilate away until after BBN, their effect is smaller.  Annihilation of \epm and the corresponding heating of photons always occur before BBN stops, and if the WIMPs are still present, the result is slower expansion at given $T_\gamma$ than if the WIMPs had already annihilated.  This decreases the size of the speed-up effect, so that at very low \mchi~(the left sides of the graphs in Fig.~\ref{fig:Deln=0yields}) the yield curves trend back towards the $\mchi\rightarrow\infty$ limit.

The helium evolution story is more complicated, as was also true for EM coupled WIMPs \cite{kngs1}.   The BBN predicted helium abundance shows effects both from the expansion timescales and from the influence of the hotter neutrinos on the weak rates.  In the EM coupled case the changes in \Yp~produced by these effects had opposite signs and partially cancelled, though they had different \mchi~dependences.  Here they all push \Yp~higher, so the deviation from the $\mchi\rightarrow\infty$ limit has the same sign at all \mchi~in Fig.~\ref{fig:Deln=0yields}.  However, neutrino-coupled WIMPs cannot shift $\eta$, so the maximum size of the shift in \Yp~is not as large for neutrino coupled as for EM coupled WIMPs.

As discussed in Ref.~\cite{kngs1}, the $^3$He and $^7$Li abundances are of less interest for the analysis presented here than are the D and $^4$He abundances.  In the case of $^3$He, this is because of the difficulty of observing primordial $^3$He directly and the apparently complicated chemical evolution  that produced the $^3$He/H observed in the Galaxy at roughly solar metallicities \cite{steig2007}.  Lithium suffers from the well-known ``BBN lithium problem":  the lithium abundance (Li/H) inferred from observations of low-metallicity stars is lower than the SBBN prediction by a factor $\gtrsim 3$~\cite{spite,fields}.  The models considered here cannot produce a downward shift of Li/H of comparable size for any choice of parameters.  Moreover, we will see below that once these parameters are fixed to reproduce the observed deuterium and helium abundances, the lithium yield is essentially fixed at its SBBN value.  Whatever is wrong with lithium, it cannot be solved by the light WIMP scenario.  A similar result was found for EM coupled WIMPs~\cite{kngs1} (and also in the absence of any light WIMP) and is driven by the anticorrelation between the BBN predicted D and Li abundances.

We now examine the results of the BBN calculations for $\Deln = 0$ (no equivalent neutrinos) and varying WIMP mass shown in Fig.~\ref{fig:wimpyields-allkinds} in the \Yp\,-- D/H plane.  The contours corresponding to the Cooke et al.~\cite{cooke} D/H constraint and the Izotov et al.~2013 \cite{izotov} \Yp~constraint are shown in Fig.\,\ref{fig:wimpyields-allkinds}, just as in Fig.~\ref{fig:bbnonly}.  The baryon density has been fixed (at the Planck $\Lambda$CDM value) and the WIMP masses varied between 40 MeV and 4 keV as in Fig.~\ref{fig:Deln=0yields}.  Starting from the SBBN limit, $\mchi\rightarrow\infty$, each curve proceeds toward higher yields of both nuclides.  There is a change in slopes near $\mchi = 3$ MeV, corresponding to changes in the slope of both yields as functions of \mchi~seen in Fig.~\ref{fig:Deln=0yields}.  Around $\mchi = 200$ keV, the decline of each curve away from its maximum in Fig.~\ref{fig:Deln=0yields} manifests itself as a sharp hook in Fig.~\ref{fig:wimpyields-allkinds}.  Finally, each curve ends at a limiting $\mchi \rightarrow 0$ point.  At this limiting point, the very light WIMP is indistinguishable (for BBN) from an equivalent neutrino: it remains relativistic all the way through BBN.  Thus, for a Majorana WIMP this final limiting point corresponds to $\mchi > 30$ MeV and $\Deln = 1$, and consequently it lies on the (short-dashed) curve of infinite \mchi~and varying \Deln.  BBN alone cannot distinguish between the parameter pairs $(\mchi > 30~\mathrm{MeV},\Deln = 1)$ and $(\mchi = 0, \Deln = 0)$ for a Majorana WIMP.  It is evident from Fig.~\ref{fig:neffvsmall} that the CMB does not share this degeneracy.  For the other WIMP types considered here, $\Deln$ is incremented not by one unit but by 4/7 (real scalar), 8/7 (complex scalar), or 2 (Dirac).  The equivalence of low \mchi~and higher \Deln~is also evident in Fig.~\ref{fig:ypvsdhneff}, considered below.

The BBN equivalence of very low \mchi~and increased \Deln, combined with the shape of the curves in Fig.~\ref{fig:wimpyields-allkinds}, implies that it is difficult for neutrino coupled light WIMPs to produce BBN yields that are very different from a model where \Deln~and \omb~are the only free parameters.  At fixed \omb, the yields with varying WIMP mass stay close to the curve of varying \Deln, with the extremes of \mchi~lying on that curve.  The farthest distance from the curve of varying \Deln~occurs at the shallow bend seen in the curves of Fig.~\ref{fig:wimpyields-allkinds}, so large changes in \mchi~are degenerate with relatively modest changes in \omb~and \Deln.

\subsection{Parameter constraints from BBN alone}

\begin{figure}[!t]
\begin{center}
\includegraphics[width=0.32\columnwidth]{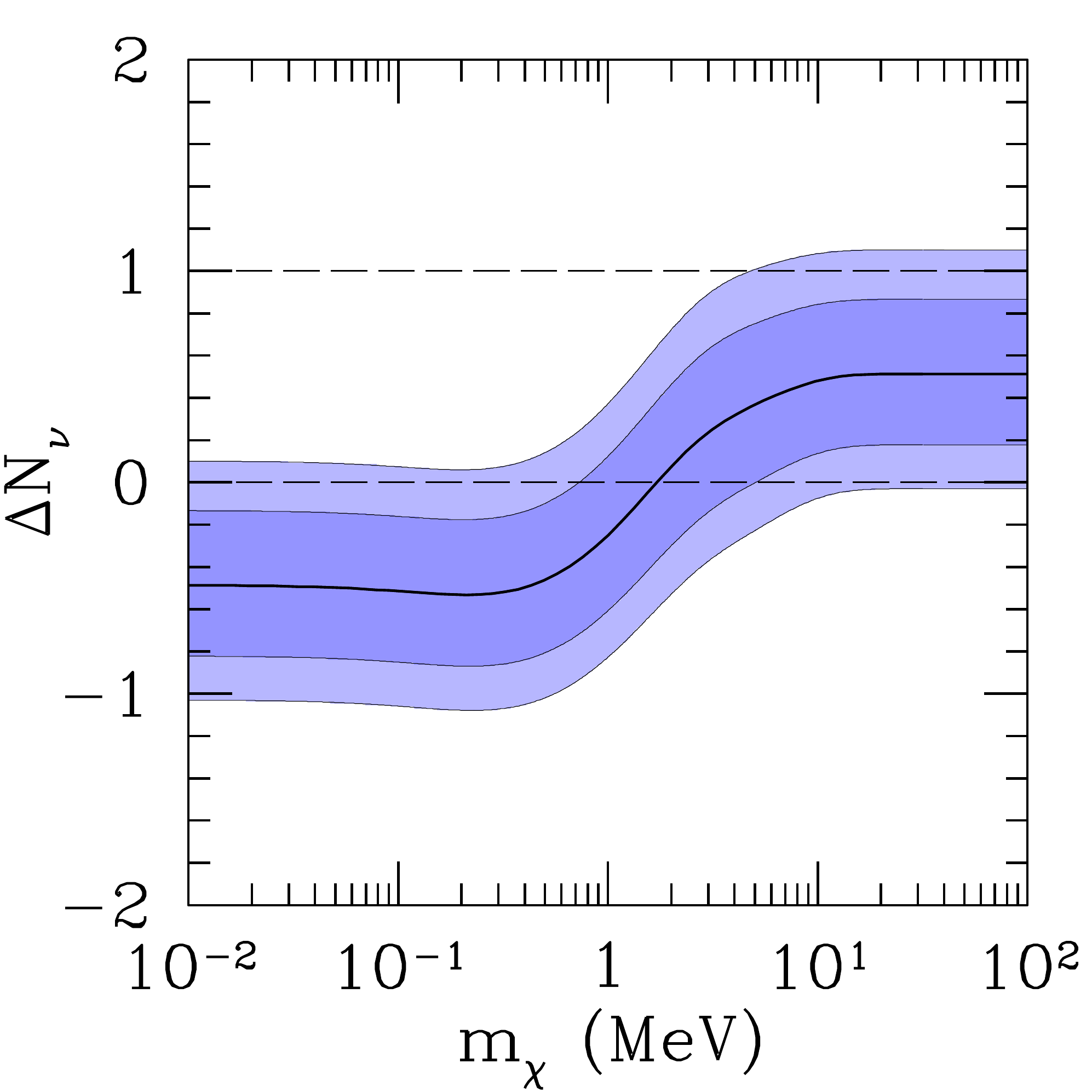}
\includegraphics[width=0.32\columnwidth]{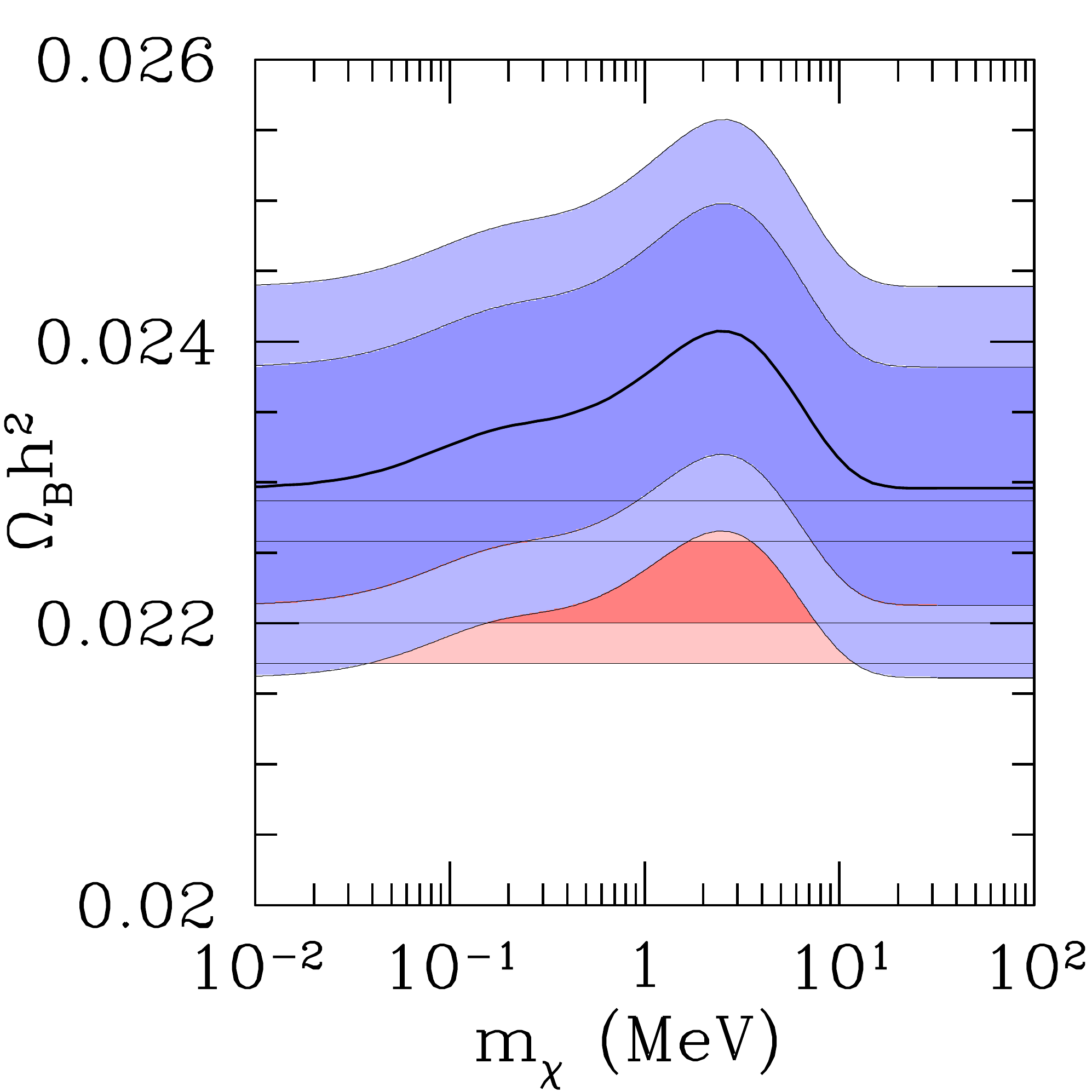}
\includegraphics[width=0.32\columnwidth]{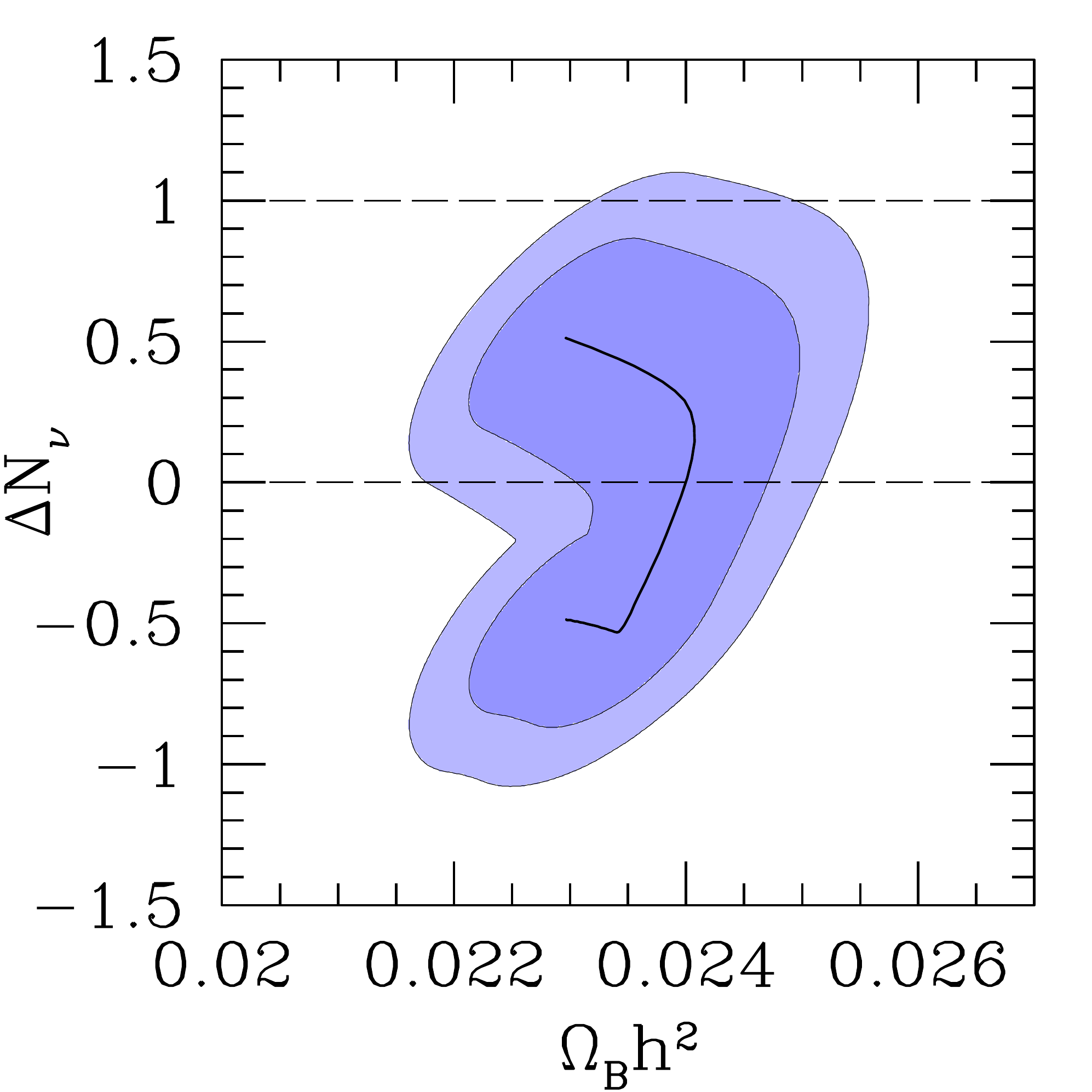}
\caption{(Color online) The left panel shows \Deln~and the middle panel shows \omb~as functions of the WIMP mass for a Majorana fermion WIMP that annihilates to SM neutrinos.  The horizontal (red/pink) bands in the middle panel show, for comparison, the CMB constraints on \omb~(that are independent of the WIMP mass).  The right panel shows the 68.3\% and 95.5\% contours for the BBN constraints in the \Deln~-- \omb~plane that follow from the left and middle panels. Darker and lighter blue (curved) contours show the BBN determined 68.3\% and 95.5\% confidence level regions of the joint likelihoods for each pair of parameters, using as constraints only D/H and Y$_{\rm P}$.  The solid (black) curve running through the middle of these regions shows the BBN best fit parameter values at each \mchi~value.  (In the right panel, this curve runs from high \mchi~at the top end of the curve -- which is the best-fit point in Fig.~\ref{fig:bbnonly} -- to low \mchi~at the bottom end.)  In the left and right hand panels the dashed lines show \Deln~= 0 and 1 as guides to the eye.   Note that the curve in the left-hand panel has the same shape (after up-down reflection) as the curve in the upper left panel of Fig.~\ref{fig:Deln=0yields}, reflecting the role of \Yp~in fitting this parameter.}
\label{fig:majorana-bbn-only}
\end{center}
\end{figure}

The three panels of Figure \ref{fig:majorana-bbn-only} show the results of employing the BBN results discussed above, for a neutrino coupled Majorana WIMP, to constrain the  cosmological parameters.  For each value of the WIMP mass there is always a pair of $\eta_{10}$ and \Deln~(or equivalently \omb~and \neff) values such that the BBN yields for D and \4he agree with their observationally inferred abundances.  This is shown by the solid (black) curves in Fig.\,\ref{fig:majorana-bbn-only}; the darker/lighter (blue) bands are 68.3\% and 95.5\% frequentist intervals in these parameter values, computed using profile likelihoods \cite{pdg2012,rolke}.  The contours shown in the left and middle panels of Fig.\,\ref{fig:majorana-bbn-only} are the BBN constraints on \Deln~and \omb~from D and \4he, as functions of the WIMP mass.  In the right hand panel of Fig.\,\ref{fig:majorana-bbn-only} the information from the left and middle panels is projected onto the $\omb - \Deln$ plane, so that \mchi~is eliminated.  The solid (black) curve in this panel shows the locus of exact fits, which vary with WIMP mass.  

The results for \Deln~as a function of \mchi~shown in the left hand panel of Fig.\,\ref{fig:majorana-bbn-only} can be understood with reference to the two upper panels of Fig.\,\ref{fig:Deln=0yields}.  As shown in \S\,\ref{sec:update1}, the observationally inferred primordial helium and deuterium abundances, \Yp~and $\yd \equiv 10^{5}(\mathrm{D/H})_{\rm P}$, can be fitted simultaneously with $\Deln \approx 0.5$ and $\omb \sim 0.022$.  Starting from this fit and reducing \mchi~from infinity ($\gtrsim 30$ MeV), the predicted values of \Yp~and \yd~both increase, so \Deln~and \omb~must be refitted.  The helium abundance is relatively insensitive to changes in the baryon density, but it is quite sensitive to changes in \Deln.  At low WIMP mass, $\mchi \lsim 10$ MeV, the BBN predicted helium abundance can be restored to consistency with the observed abundance by reducing \Deln.  At very low \mchi, a Majorana WIMP affects \Yp~in exactly the same way as a sterile neutrino, so the best fit at very low mass is one unit of \Deln~lower than the best fit for $\mchi > 30$ MeV.   In fact, the curve of \Deln~vs.~\mchi~in Fig.~\ref{fig:majorana-bbn-only} is very nearly the curve of \Yp~vs.~\mchi~in Fig.~\ref{fig:Deln=0yields}, flipped upside down.  

Lowering \Deln~also lowers \yd, helping to reconcile it with a light WIMP, but the \Deln~that fits \Yp~is not low enough to bring \yd~back into full agreement.   Deuterium is more sensitive than \Yp~to \omb, which can be increased to further reduce \yd~into agreement with the data.   This is visible in the middle panel of Fig.\,\ref{fig:majorana-bbn-only} in the mass range $\sim 1 - 10$ MeV.  Just as for \Yp, low WIMP masses ($\lsim 0.1$ MeV) can be compensated in \yd~exactly by a unit decrease of \Deln, so the \omb~that provided the exact fit at high \mchi~is also the best fit at low \mchi.  This is also visible in the middle panel of Fig.\,\ref{fig:majorana-bbn-only}.

In carrying out our calculations, we have permitted $\Deln < 0$ by allowing for negative values of the equivalent neutrino energy density (though the sum of SM and equivalent neutrino energy densities remains positive).  Especially in the BBN-only analysis, this is allowed formally by the observational data.  However, since there are known to be three standard-model neutrino species, $\Deln < 0$ cannot be achieved by adding new species as contemplated here, and it is unphysical without further assumptions.  There are two ways to address this issue.  We could impose directly an \textit{a priori} constraint that $\Deln \geq 0$.   Alternatively, we could use the CMB (Planck) constraint on \neff~\cite{planck} and the assumption that $\Deln \geq 0$ to set a lower limit on \mchi, and then fit the parameters without imposing a sharp limit on \Deln.  We have done both and find that the parameter constraints in the two approaches are essentially equivalent (unless the WIMP is a real scalar, which affects the CMB too little for this approach to work).  Next, in \S\,\ref{sec:joint}, we describe the results of the joint CMB + BBN analysis. The merits of imposing the cut on \mchi~instead of directly on \Deln~will be evident there.

\begin{figure}
\includegraphics[width=0.5\columnwidth]{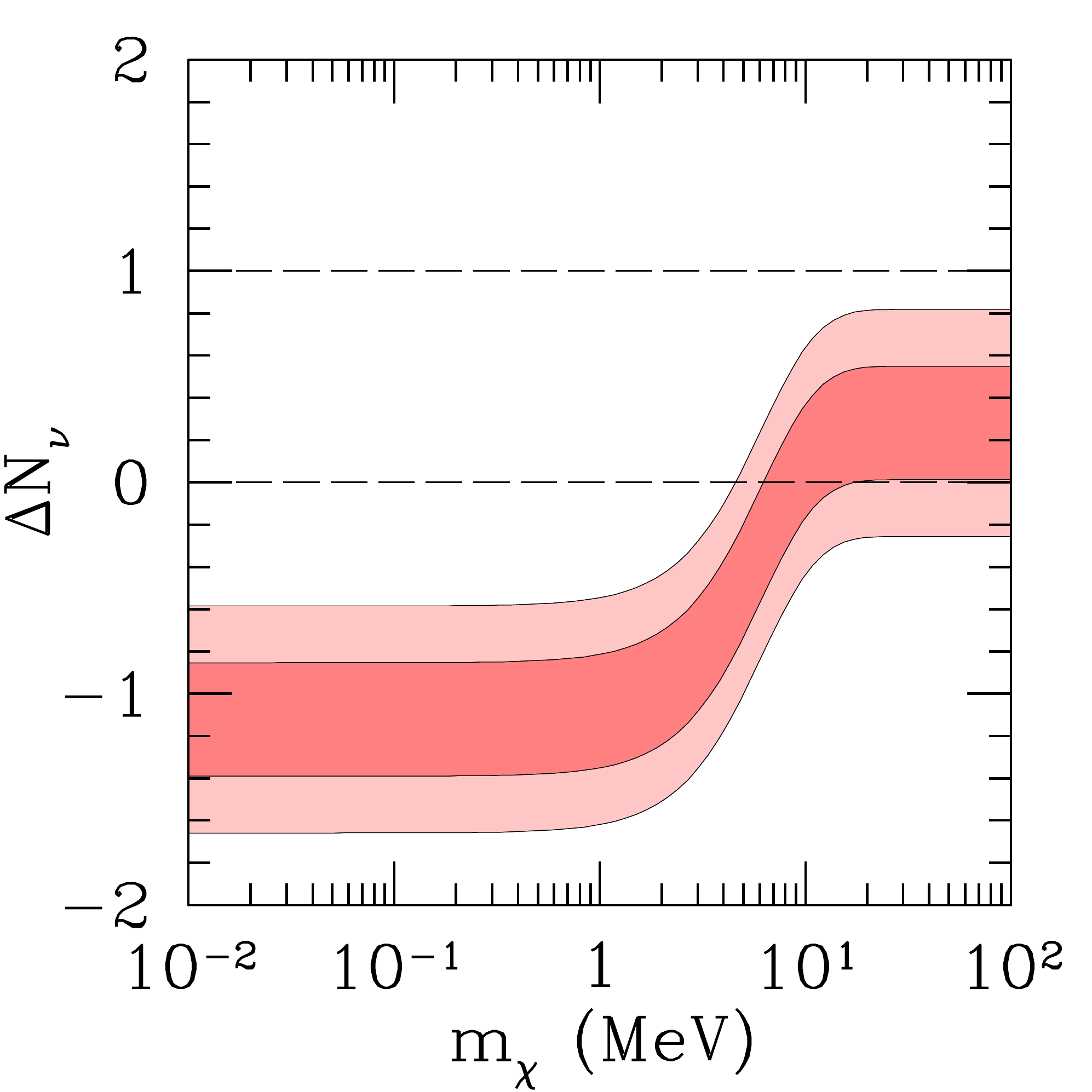}
  \caption{(Color online)  \Deln~as a function of the (Majorana) WIMP mass, showing 68.3\% and 95.5\% confidence limits inferred from the Planck CMB data (Eq.~74 of Ref.~\cite{planck}) and Eq.~\ref{eq:neff-nnu-caseA}
 ($\Deln \approx \neff({\rm CMB}) - \neff^{0}(\mchi)$).}
  \label{fig:delncmb}
\end{figure}

\begin{figure}[!t]
\begin{center}
\includegraphics[width=0.32\columnwidth]{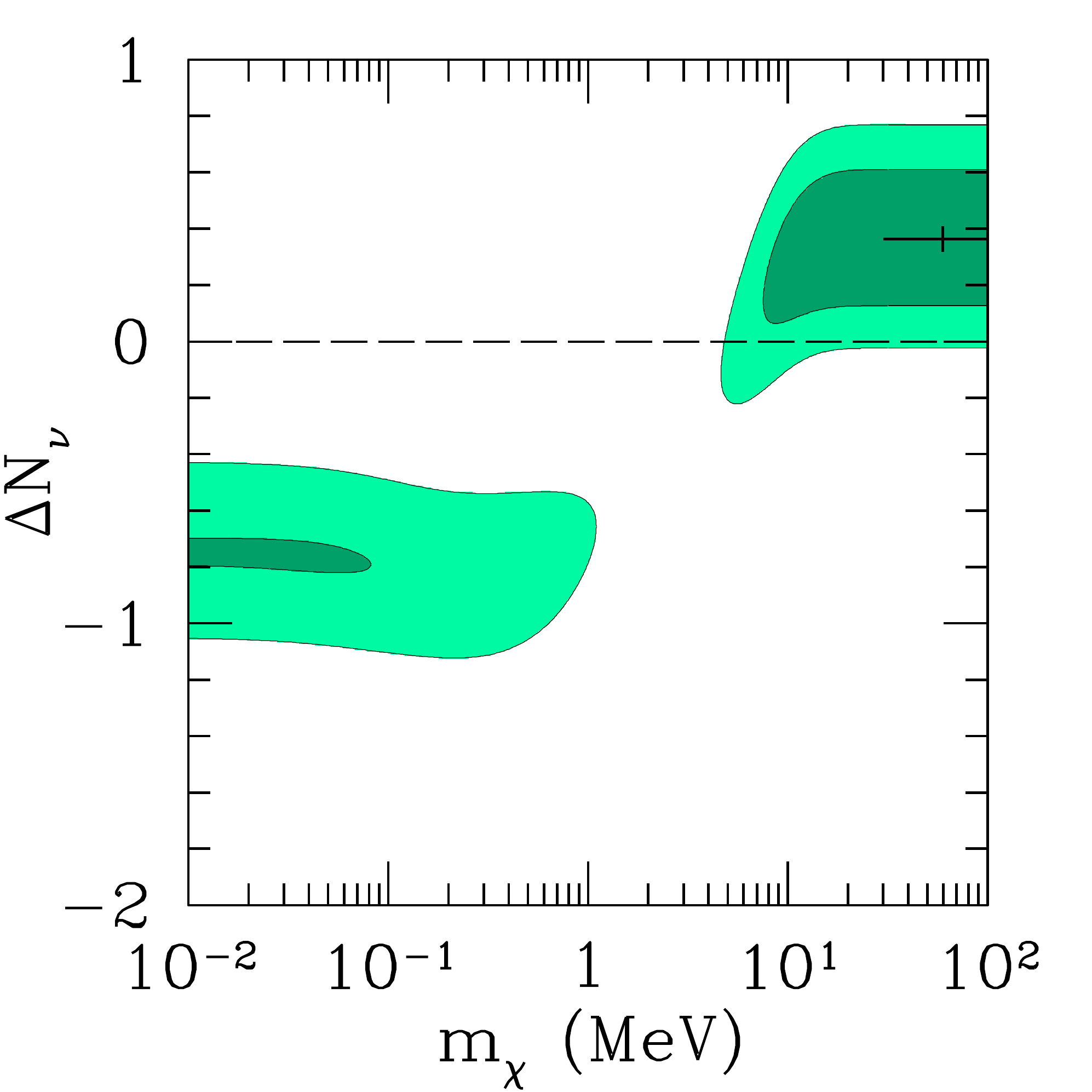}
\includegraphics[width=0.32\columnwidth]{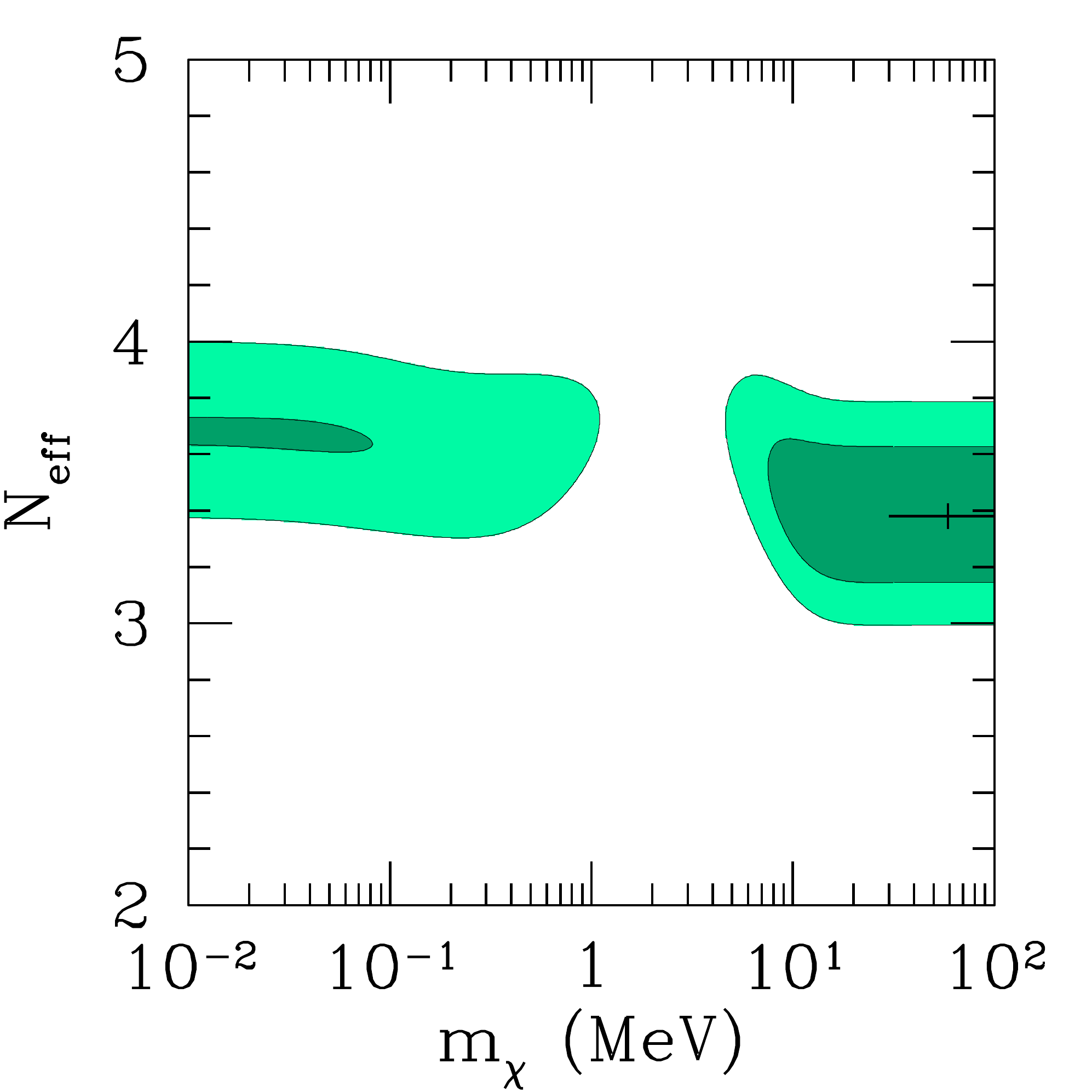}
\includegraphics[width=0.32\columnwidth]{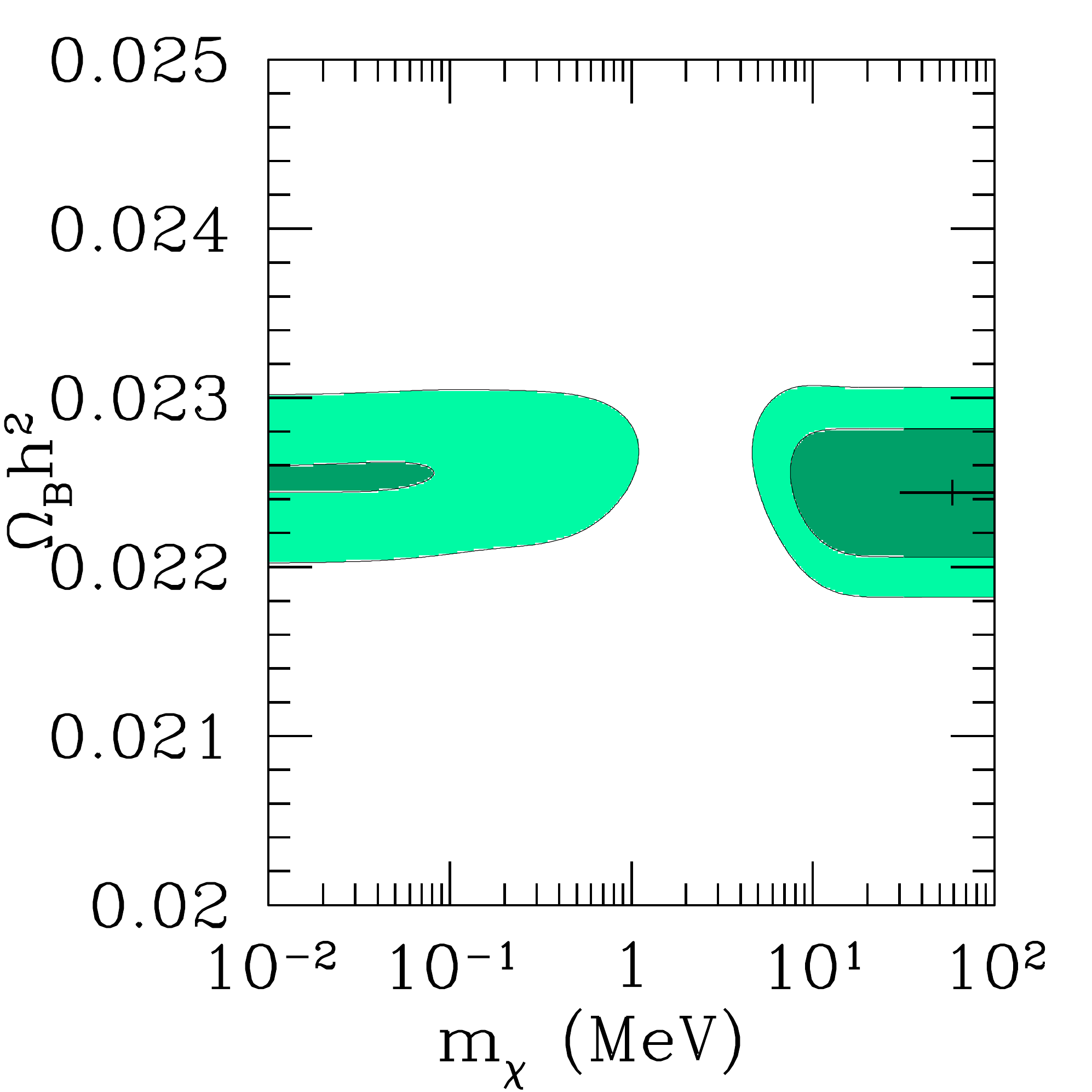}
\caption{(Color online) The joint BBN + CMB two-parameter 68.3\% and 95.5\% confidence contours of \mchi~and \Deln~(left panel), \neff~(middle panel), and \omb~(right panel).  Results are shown for a Majorana WIMP that annihilates to SM neutrinos.  The ``+" shows the best fit point in the parameter space, which is within $\Delta\chi^2 = 10^{-4}$ of $\mchi \rightarrow\infty$.  In the left hand panel the dashed line shows \Deln~= 0  as a guide to the eye.  Since these contours indicate the joint constraints on two parameters, they correspond to different $\chi^2$ values than would the boundaries on the single parameters in Eqs.~(\ref{eq:nucoupled-constraints}) and (\ref{eq:nucoupled-constraints-2s}).}
\label{fig:majorana-bbn-cmb}
\end{center}
\end{figure}

\begin{figure}[!t]
\begin{center}
\includegraphics[width=0.45\columnwidth]{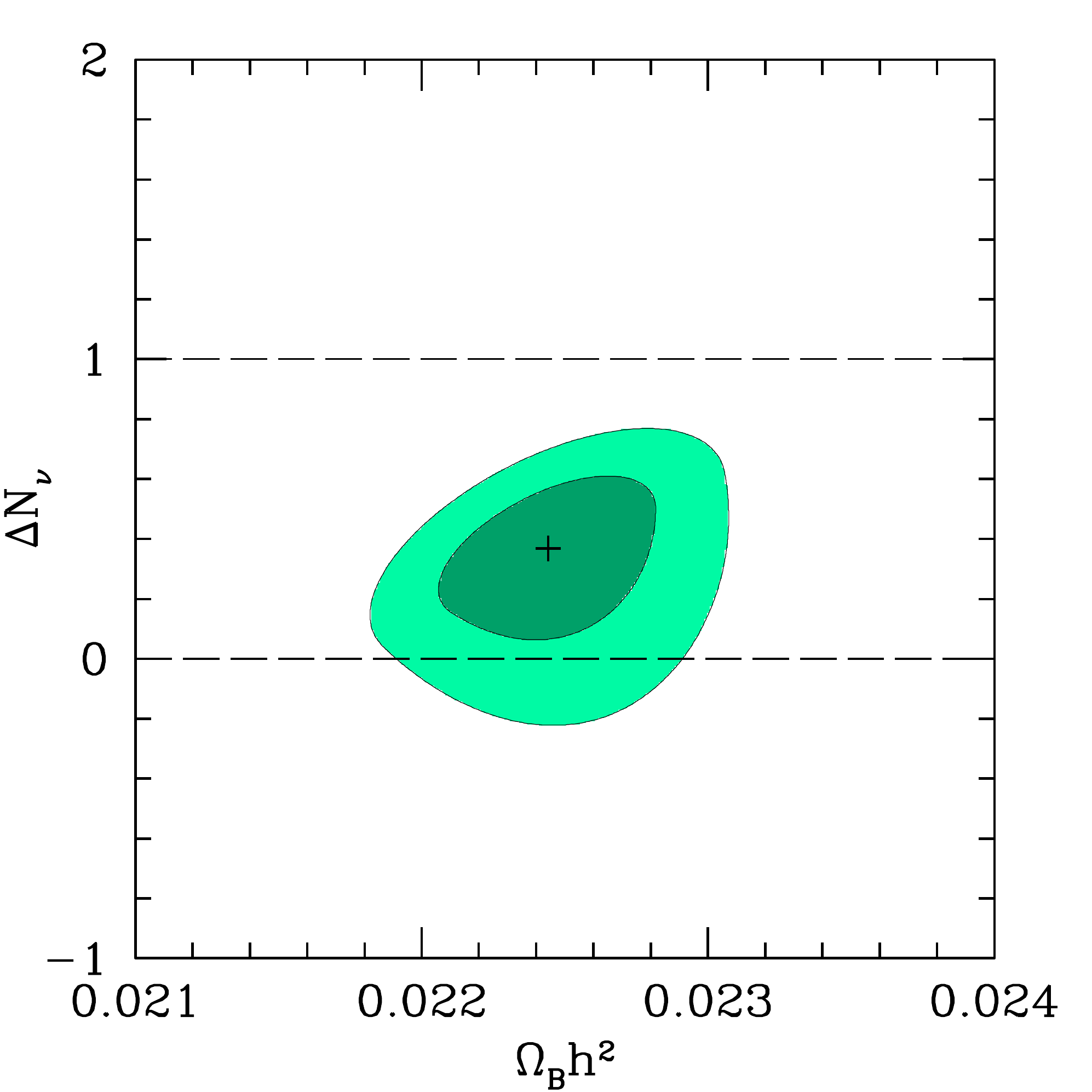}
\includegraphics[width=0.45\columnwidth]{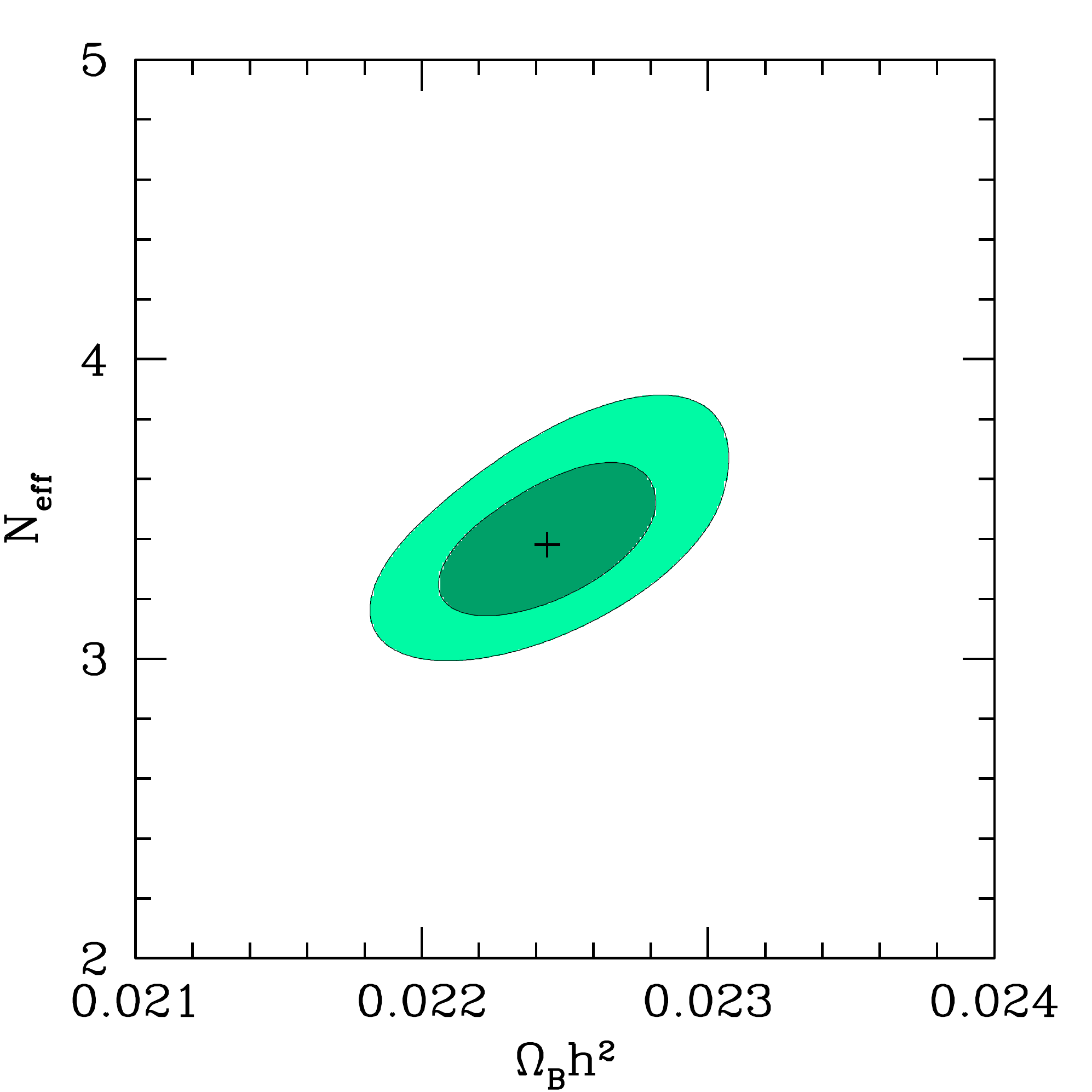}
\caption{(Color online)  Two-parameter joint contours of \Deln~(left panel) and \neff~(right panel) with the baryon mass density (\omb)  for a Majorana fermion WIMP that annihilates to SM neutrinos.  Limits were determined by  joint BBN + CMB fitting with the restriction that $\mchi \gsim 2$ MeV (see the text).  The ``+'' shows the best fit point in each panel.  In the left hand panel the dashed lines show \Deln~= 0 and 1 as guides to the eye.}
\label{fig:majorana-bbn-cmb-highmchi}
\end{center}
\end{figure}

\section{Joint BBN And CMB Constraints On A Neutrino Coupled WIMP}
\label{sec:joint}

The upper set of curves in Fig.\,\ref{fig:neff0vsmall} shows, for neutrino coupled WIMPs, $\neff^{0}$ as a function of the WIMP mass, where $\neff^{0}$ is the value of \neff~when $\Deln = 0$.  In comparison with the Planck constraint on \neff\,\cite{planck}, it is clear that  at low WIMP mass, consistency with the CMB requires $\Deln < 0$ (though the requirement is weak if the WIMP is a real scalar).  This is shown in Figure \ref{fig:delncmb}, where the fitted $\Deln \approx \neff(\rm Planck) - \neff^{0}$ (cf.~Eq.~\ref{eq:neff-nnu-caseA}) is shown as a function of the WIMP mass, for a Majorana fermion WIMP.  Comparing this result with those of the previous section, we see that in the low mass limit, both the CMB and BBN require $\Deln < 0$ to accommodate very light neutrino coupled WIMPs.  The joint BBN and CMB constraints can be used to separate the unphysical, $\Deln < 0$ mass range, from the mass range where $\Deln \gtrsim 0$.  This is illustrated for Majorana WIMPs in Figure \ref{fig:majorana-bbn-cmb}, where the left hand panel shows the joint BBN + CMB best fit and 68.3\% and 95.5\% simultaneous constraints on \Deln~and \mchi, and the right hand panel shows simultaneous constraints on \omb~and \mchi.  (Likelihoods involving the CMB constraints are again computed as in \S\,\ref{sec:nowimp}.)  As is revealed by these figures, there are two ``peninsulas'' of acceptable fits to the combined BBN and CMB data, separated by a region of very low likelihood.  One is located at relatively high WIMP mass ($\mchi \gsim 5$ MeV)  corresponding to $\Deln \gtrsim 0$ and with minimum $\chi^2 \sim 1.2$ (1 degree of freedom), while the other is at low WIMP mass ($\mchi \lsim 1$ MeV), corresponding to $\Deln < 0$ and larger minimum $\chi^2$ (1.8 for real scalars, 3.3 for Majorana fermions, and 10.2 for Dirac fermions).   Examination of Figs.~\ref{fig:delncmb} and \ref{fig:majorana-bbn-cmb} indicates that the physical regime where $\Deln \geq 0$ occurs only where $\mchi \gsim 2$ MeV, so we restrict our analysis to $\mchi > 2$ MeV in order to isolate the high-\mchi~peninsula in the fitting, unless otherwise stated.  Having ruled out the ``pensinsula'' at low \mchi~in this way, we arrive at the simultaneous constraints for \Deln, \neff, and baryon density (\omb) shown in Fig.~\ref{fig:majorana-bbn-cmb-highmchi}, and the lower limits on \mchi~shown in Table \ref{tab:masstable}.

The best fit point is at $\mchi \gtrsim 35$ MeV, which is in the $\Deln \geq 0$ regime, and is essentially at the limit of no light WIMP\,\footnote{For all cases except for real scalar WIMPs, a formal best fit was found at $\mchi \approx 35$ MeV, though this is only better than the $\mchi \rightarrow \infty$ limit by $\Delta \chi^2 \sim 10^{-4}$.  This corresponds to tiny differences in the computed quantities, smaller than the precision of both the data constraints and the underlying inputs such as nuclear rates; it is comparable to the precision of the interpolation table used to provide BBN outputs for the likelihood calculation.}.  For all types of neutrino coupled WIMPs, the preferred parameter values and their $1\,\sigma$ ranges are very close to those in Eq.~(\ref{eq:bbn+cmb-nowimp-fit}) for no light WIMP at all:
\begin{eqnarray}\label{eq:nucoupled-constraints}
 \Deln & = & 0.37^{+0.16}_{-0.17} \\\nonumber
\neff & = & 3.38^{+0.17}_{-0.16} \\\nonumber
100\,\omb & = & 2.24 \pm 0.03 
\end{eqnarray}
The difference in the best-fit parameters between these results and those in Eq.~(\ref{eq:bbn+cmb-nowimp-fit}) lies entirely in the corrections to the IND approximation included in the earlier analysis; we did not attempt such a correction when including light WIMPs.  In Fig.~\ref{fig:majorana-bbn-cmb-highmchi}, the contours can be seen as simple ellipses centered on the high-\mchi~best fit from Fig.~\ref{fig:bbncmb}, superimposed with regions allowed by the dip in \Deln~at $\mchi \lesssim 10$ MeV in Fig.~\ref{fig:majorana-bbn-cmb}.  This allowed region of slightly lowered \mchi~introduces the asymmetry seen at $1\,\sigma$ in Eq.~(\ref{eq:nucoupled-constraints}), rendering the projection of the likelihoods onto the \Deln~axis non-Gaussian; multiples of the $1\,\sigma$ error are poor guides to the tails of the distribution.  In particular, having isolated high-\mchi~contours only by restricting the WIMP mass to $\mchi > 2$ MeV, we still obtain a small region allowed at 95.5\% confidence where $\Deln < 0$.  Boundaries of 95.5\% confidence in single parameters (ignoring \omb, which
does not deviate significantly from Gaussian likelihood) are
\begin{eqnarray}\label{eq:nucoupled-constraints-2s}
 \Deln & = & 0.37^{+0.32}_{-0.44} \\\nonumber
\neff & = & 3.38^{+0.38}_{-0.31}\,.
\end{eqnarray}

For a real scalar WIMP, there is no way to isolate the high-\mchi~likelihood minimum at a level that matters for the limits presented here.  This is because it is possible to traverse the region between the ``peninsulas'' in Fig.~\ref{fig:majorana-bbn-cmb} without passing through a point where $\Delta \chi^2 \geq 4$ (relative to the global best fit), and also because the local minimum of $\chi^2$ in the low-\mchi~regime has $\Delta \chi^2 < 1$.  Requiring that $\Deln \geq 0$ for real scalars, it is possible to obtain a weak lower limit on \mchi~(Table \ref{tab:masstable}), and the 95.5\% (two-sided) upper limit on \neff~becomes $3.83$.  Limits are otherwise the same as in the other cases.  As in the absence of a light WIMP (but not the presence an EM coupled WIMP), a sterile neutrino (\Deln~= 1) is strongly disfavored, at more than 99\% confidence, in the presence of a neutrino coupled WIMP.  However, the complete absence of equivalent neutrinos is also disfavored in all cases, but ``only'' at the $\sim 98\%$ confidence level.

As described in \S\,\ref{sec:nureview} and \S\,\ref{sec:bbn-calc}, there exist the two possibilities that the light WIMP couples only to the SM neutrinos (resulting in Eq.~(\ref{eq:neff-nnu-caseA})) and that it couples to both SM and equivalent neutrinos (resulting in Eq.~\ref{eq:neff-nnu-caseC})).  The results quoted here were all computed for coupling only to SM neutrinos, which would seem to be the physically more interesting case.  Within the precision of rounding, the same results as Eqs.~(\ref{eq:nucoupled-constraints}) and (\ref{eq:nucoupled-constraints-2s}) are found in the other case, provided that \Deln~is still permitted to vary continuously, and mass limits within 200 keV of those in Table \ref{tab:masstable} result.  The two sets of results are guaranteed to be close by the location of the best fit at $\mchi \rightarrow \infty$, the small allowed values of \Deln, and the passage of the BBN-only best fit curve (Fig.~\ref{fig:majorana-bbn-only}) through $\Deln = 0$; as discussed in \S\,\ref{sec:bbn-calc} above, there is no difference between the two possibilities when $\mchi \rightarrow \infty$ or $\Deln = 0$, so results for the two cases never get very far apart.

\begin{table}
  \caption{Joint BBN+CMB lower limits on the mass of a neutrino coupled WIMP for WIMPs with the different spin degeneracies indicated.   These are one-sided 95.5\% limits, so that 4.5\% of the likelihood lies at lower \mchi.  The CMB constraints on \omb~and \neff~in the first line are from the Planck $\Lambda\mathrm{CDM}+N_\mathrm{eff}$ fit including BAO (Eq.~74 of Ref.~\cite{planck}).  In the second line, the Planck fit includes \Yp~as a fitted parameter (Eq.~89 of Ref.~\cite{planck}), and this constraint on \Yp~is included in our fit.  The correlations between Planck parameters have been included as described in Ref.~\cite{kngs1}.}
\begin{tabular}{lcccc}
Inputs & \multicolumn{4}{c}{Minimum \mchi~(MeV)}\\   
       &  Real scalar & Majorana & Complex scalar & Dirac\\
\hline\hline
BBN+Planck $\Lambda\mathrm{CDM}+N_\mathrm{eff}$ & 4.16 & 6.93 & 6.98 & 9.28 \\
BBN+Planck $\Lambda\mathrm{CDM}+N_\mathrm{eff}+Y_P$ & --\footnote{No value of \mchi~is ruled out at 95.5\% significance.} & 6.68 & 6.74 & 9.07\\ 
\end{tabular}
  \label{tab:masstable}
\end{table}

\subsection{Lower Bound To The Mass Of A Neutrino Coupled Light WIMP}
\label{sec:lower bound}

The lower bound to the mass of a neutrino coupled WIMP is of interest for limits on possible dark matter masses and elastic cross sections \cite{beacom,palomares,deppisch}.  The lower bounds to the mass of a neutrino coupled light WIMP derived from our joint BBN+CMB analysis are presented in Table\,\ref{tab:masstable}.  It is important to consider whether the limits on \mchi~in Table \ref{tab:masstable} truly rule out all lower masses, or if there is some mass below which our analysis no longer applies.  In the BBN calculation it is assumed that the light WIMP remains coupled to the SM neutrinos during BBN, but not necessarily after BBN has ended.  If this were not true, and WIMPs decoupled when still relativistic during or before BBN, the relic number density of light WIMPs today would be similar to that of the SM neutrinos.  In this case, any \mchi~greater than a few eV would contribute too much to the cold, warm, or hot mass density in the present universe.  Thus, any WIMP massive enough to annihilate during or before BBN must have annihilated then, and any lighter WIMP remains coupled to the SM neutrinos all the way through BBN, as assumed in the analysis here.  From Fig.~\ref{fig:majorana-bbn-only} and analogous calculations for other WIMP types, BBN alone, along with the requirement that $\Deln \geq 0$, forbids all $\mchi < 600$ keV (800 keV, 2.1 MeV) for a Majorana WIMP (complex scalar WIMP, Dirac WIMP), without exception.  At the 95.5\% confidence level, the BBN constraint alone does not forbid a real scalar WIMP of any mass.

For the joint BBN+CMB limit on \mchi~from the analysis presented here to be valid, it is also necessary that the relation between \mchi~and $\neff^0$ shown in Fig.~\ref{fig:neff0vsmall} holds.  This requires that the light WIMP remain in equilibrium with the SM neutrinos until after it has annihilated (at $T_\nu \sim \mchi/3$), in order that $n_{\chi}/n_{\nu} \ll 1$ well before recombination.  Thus, as mentioned above, any neutrino coupled WIMP with mass greater than a few eV must have annihilated prior to recombination (or, prior to the epoch of equal matter-radiation densities), heating the SM neutrinos as assumed in the joint BBN+CMB analysis here.  

A light WIMP that is still relativistic after recombination ($\mchi \lesssim 0.1\,{\rm eV}$) should count in the CMB as a contribution to \Deln, not as a source of higher $T_\nu$.  Thus, at extremely low masses, the light WIMP meets our definition of an equivalent neutrino; if the upper four curves of Fig.~\ref{fig:neff0vsmall} were continued below $\sim 0.1$ eV, they would eventually drop to $\neff = 3.02(1+\Delta N_{\nu,\chi}/3)$, where $\Delta N_{\nu,\chi} = 4/7,\,1,\,8/7,$ or 2 for the particle types considered here.  With the exception of the real scalar, an extremely light WIMP is a bad fit to the CMB data alone, although detailed quantitative analysis of $\mchi \lesssim {\rm few}$ eV would require dedicated modeling of CMB power spectra.

\subsection{Primordial Abundance of \4he}

It should be noted that empirical values of the primordial helium abundance exist in the literature that differ from the value of \Yp~we have adopted from Izotov et al.~\cite{izotov}.  In particular, Aver et al.~\cite{aver13} reanalyzed a subset of the Izotov data by different methods and rejected data for individual H{\sc ii} regions that they considered to be poorly modeled.  For the mean of the helium abundances derived from the small remaining data subset, they find $0.2535\pm 0.0036$, entirely consistent with the value \Yp~$= 0.254\pm 0.003$ used here.  Following the usual assumption that the post-BBN helium abundance should correlate with the metallicity (\eg, the oxygen abundance), Aver et al.~fitted a linear regression to their helium abundance versus metallicity data and reported its intercept as \Yp.  The small amount of data included in the Aver et al.~final sample, and the small metallicity range they cover, allow a wide range of regression slopes that propagates into
large errors on the inferred intercept, \Yp~$= 0.2465\pm 0.0097$.  Within this wide range, the Aver et al.~estimate overlaps with the Izotov value used here, but it provides poor parameter constraints. Given the limited data set and the limited metallicity range, in our opinion it would be a better use of these data to avoid a metallicity correction by considering the mean (perhaps restricted to the very lowest-metallicity H{\sc ii} regions) than to introduce an avoidable large error in the correction.  Using the large Aver et al.~error in our analysis allows a very wide range of \Deln~from BBN (roughly $-2$ to $+1.3$ at 95\% confidence if there is a Majorana light WIMP, and roughly $-1.5$ to $+1.3$ if not), along with any value at all of \mchi, and it results in joint BBN+CMB constraints that are reduced nearly to those obtained from the CMB alone.

\section{Summary And Conclusions}
\label{sec:summary}

In the absence of a light WIMP and equivalent neutrinos ($\Deln \equiv 0$), BBN (in this case, SBBN) depends only on the baryon density.  As shown in \S\,\ref{sec:update1}, the SBBN-predicted abundances of deuterium and helium are a poor fit ($\chi^{2}_\mathrm{min} = 5.3$) to the observationally-inferred primordial abundances \cite{cooke,izotov}, although the baryon density of this fit is in excellent agreement with the value inferred, independently, from the Planck CMB data \cite{planck}.  When \Deln~is also fitted simultaneously (still without a light WIMP), an exact fit ($\chi^{2} = 0$) of the two observed abundances results.  Accounting for the observational uncertainties in the primordial abundances and correcting for late neutrino heating, BBN alone gives $100\,\omb = 2.29 \pm 0.06$, $\Deln = 0.50 \pm 0.23$, and $\neff = 3.56 \pm 0.23$ (see \S\,\ref{sec:update1} and Fig.\,\ref{fig:bbnonly}).  These parameter values are in excellent agreement with those determined independently from the CMB \cite{planck}, and a joint BBN + CMB fit gives $100\,\omb = 2.24 \pm 0.03$, $\Deln = 0.35 \pm 0.16$ and, $\neff = 3.40 \pm 0.16$ (see Fig.\,\ref{fig:bbncmb}).  While $\Deln \leq 0$ is excluded at $\sim 98\%$ confidence (i.e., 98\% of the likelihood lies at $\Deln > 0$), $\Deln \geq 1$ is also excluded at $\gsim 99\%$ confidence\,\footnote{After the work described here was finished, we became aware of a preprint presenting a new estimate of the primordial helium abundance, \Yp~$= 0.2551 \pm 0.0022$ \cite{izotov14}.  This small change in the value of \Yp~from that adopted here would, on its own, increase the BBN-inferred value of \Deln~to $\Deln = 0.59 \pm 0.17$ (while reducing the error), still disfavoring both \Deln~= 0 and \Deln~= 1.  Simultaneous fitting of this result with Planck parameter constraints yields $\Deln = 0.45\pm 0.13$.  These new parameters, resulting from the revised \Yp~value (and its error), are entirely consistent with the previous result within the errors of the inputs.  The shift noted here would be largely canceled if the systematic difficulties with the neutron lifetime were to be resolved in favor of the beam measurements \cite{wietfeldt-greene}.}.

In \S\,\ref{sec:update2} the case of an electromagnetically coupled light WIMP \cite{kngs1} was revisited, adopting the revised primordial D abundance \cite{cooke}.  The parameter estimates are changed very little from those presented in Ref.~\cite{kngs1}.  The BBN + CMB best fit remains $\Deln = 0.65^{+0.45}_{-0.37}$ and (in the IND approximation, where the constant in Eqs.~(\ref{eq:neff-nnu-caseA}) and (\ref{eq:neff-nnu-caseC}) is 3.02) $\neff = 3.22 \pm 0.25$.  The lower limits on \mchi~are about 100 keV higher than those shown in Ref.~\cite{kngs1}.  Given an electromagnetically coupled light WIMP with \mchi~in the allowed range, a sterile neutrino (\Deln~= 1) is still allowed at $1\sigma$, while \Deln~= 0 is still disfavored.

The main focus of this paper has been the consideration of the effects on BBN (and the CMB) of a light WIMP that only couples to (and, until $T_\gamma \ll m_\chi$, remains in chemical equilibrium with) neutrinos (SM and/or equivalent; \S\,\ref{sec:bbnnuwimp} and \S\,\ref{sec:joint}).  We find that, well within the errors, the results for annihilation to SM neutrinos and to SM and equivalent neutrinos together are indistinguishable, and we have presented quantitative results for a WIMP that annihilates to the SM neutrinos.  The key difference between this case and that of a light WIMP that annihilates to photons and/or \epm pairs is that here, as \mchi~decreases, $\neff^{0}$ increases from $\neff^{0} \approx 3$, instead of increasing.  To maintain consistency with the Planck value of $\neff \approx \neff^{0} + \Deln = 3.30 \pm 0.27$, requires an unphysical value of $\Deln < 0$ for sufficiently small WIMP masses (see Figs.\,\ref{fig:neff0vsmall} and \ref{fig:neffvsmall}).  The increase of D and \4he yields at low $m_\chi$ (Fig.~\ref{fig:Deln=0yields}) similarly requires $\Deln < 0$ to maintain agreement with observations.  Limiting \Deln~to non-negative values sets a lower bound to the WIMP mass, allowing only values in the regime where the WIMP has little effect on BBN (unless the WIMP is a real scalar, for which $m_\chi \ll m_e$ is not strongly ruled out).  For all neutrino coupled WIMPs, the best fit WIMP mass is $\mchi \geq 35$ MeV, and joint BBN + CMB fits yield $100\,\omb = 2.24 \pm 0.03$, $\Deln = 0.37^{+0.16}_{-0.17}$, and (again in the IND approximation), $\neff = 3.38 \pm 0.16$, in excellent agreement with the parameter values inferred without a light WIMP.  As for the case of no light WIMP, for a neutrino coupled WIMP, \Deln~= 0 is disfavored (at $\sim 98\%$ confidence) and a sterile neutrino is excluded at $\gsim 99\%$ confidence.

\begin{figure}[!t]
\includegraphics[width=0.45\columnwidth]{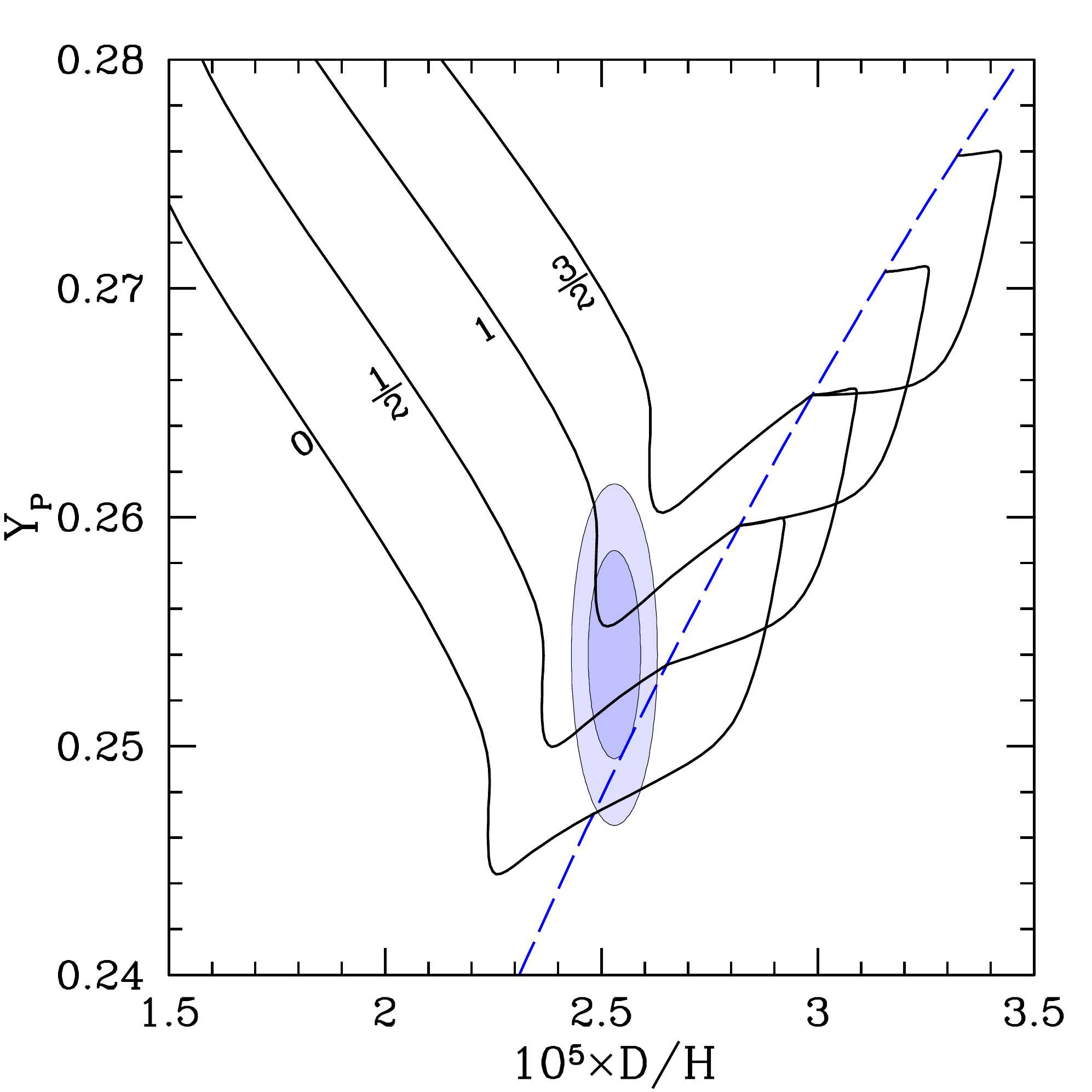}
\includegraphics[width=0.45\columnwidth]{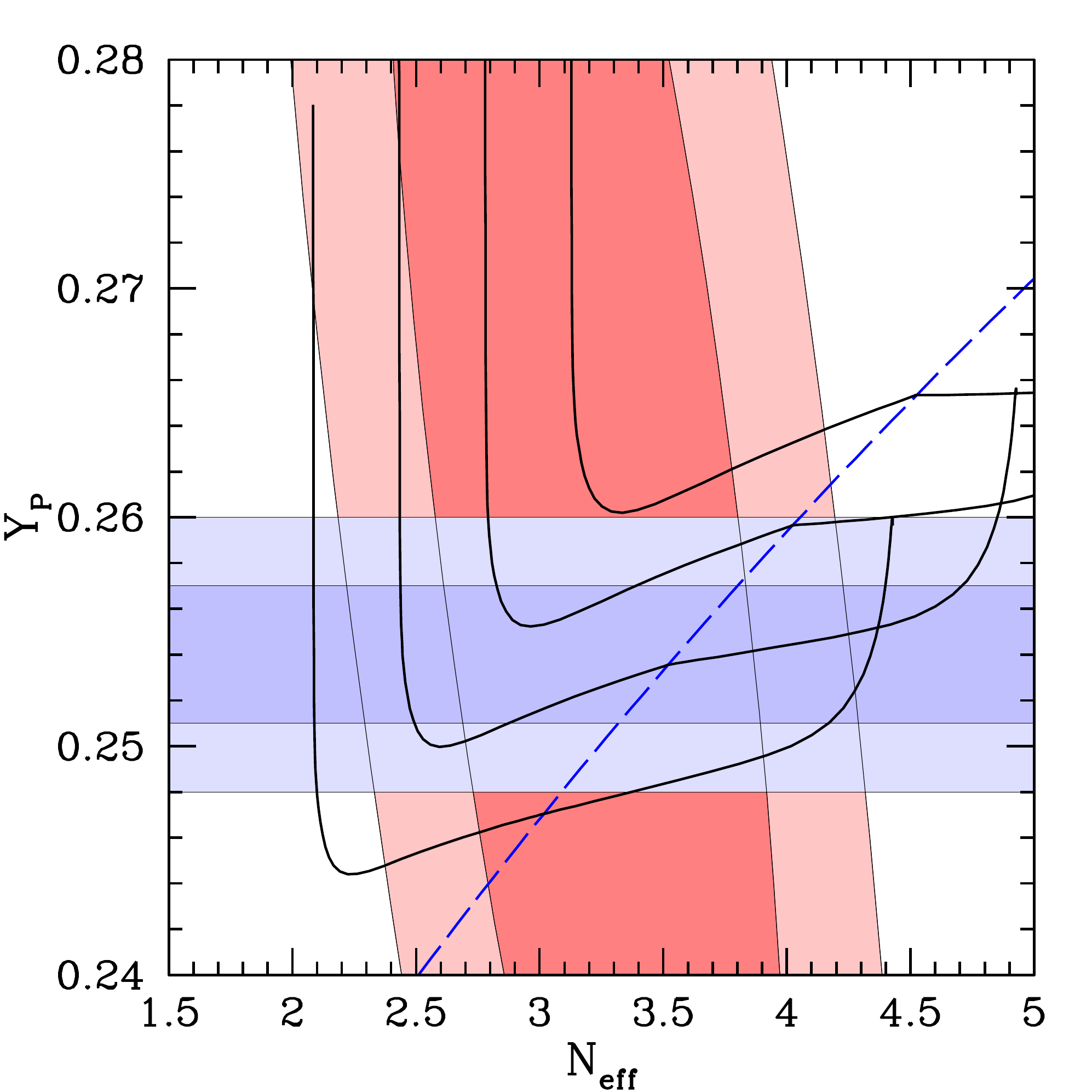}
\caption{(Color online)  The left hand panel combines the results for \Yp~versus D/H shown in Fig.\,\ref{fig:wimpyields-allkinds} for the neutrino coupled WIMP, with the analogous Figure 5 of Ref.~\cite{kngs1} for the case of an electromagnetically coupled WIMP (all at fixed \omb~from Planck).  The solid (black) curves show the BBN predicted helium and deuterium abundances for several values of \Deln~as indicated.  The dashed (blue) curve showing the $\mchi \rightarrow \infty$ limit separates electromagnetically coupled WIMPs (on the left) from neutrino coupled WIMPs (on the right).  At the upper left ends of the curves, the WIMP mass is very small.  As \mchi~increases, the BBN predicted abundances move along the curves, reaching the dashed curve in the high WIMP mass limit.  Then, proceeding on the right side of the dashed curve, the WIMP mass decreases.  The shaded contours show joint 68.3\% and 95.5\% contours for the observationally inferred, primordial deuterium \cite{cooke} and helium \cite{izotov} abundances.  The right hand panel shows the analogous results for \Yp~versus \neff.  The solid (black) curves and the solid (blue) curve correspond to those in the left hand panel.  The horizontal bands (blue) are for the 68.3\% and 95.5\% helium abundance contours and the contours (pink) from the upper left to the lower right are from the Planck \neff~and \Yp~fit (Ref.~\cite{planck} and Table II of Ref.~\cite{kngs1}).}
\label{fig:ypvsdhneff}
\end{figure}

The results presented here for neutrino coupled WIMPs and in Ref.~\cite{kngs1} for electromagnetically coupled WIMPs may be understood in a single framework with reference to the two panels of Figure \ref{fig:ypvsdhneff}, which show results for Majorana WIMPs.  The left hand panel, for BBN alone, shows the observationally inferred 68.3\% and 95.5\% confidence contours of \Yp~\cite{izotov} and D/H \cite{cooke} adopted for the analysis here.  The solid black curves show the BBN \Yp~and D/H predictions at fixed (Planck $\Lambda$CDM) \omb~for WIMPs that couple electromagnetically or only to neutrinos, for four possible \Deln~values.  At the upper left end of each curve, we have electromagnetically coupled WIMPs with very small \mchi.  Descending along the curve, \mchi~increases, \Yp~decreases and D/H increases, eventually reaching the dashed blue curve where $\mchi \rightarrow \infty$ (equivalent to no light WIMP).  To the right of the dashed curve the WIMP is coupled to standard model neutrinos.  Continuing along the solid curve, \mchi~decreases, ending in the $\mchi \rightarrow 0$ limit.  For more details about this behavior, see \cite{kngs1} and \S\,\ref{sec:bbnnuwimp}.  As may be seen from the left hand panel of Fig.\,\ref{fig:ypvsdhneff}, BBN in the presence of a sufficiently light ($\mchi \approx 5 - 10$ MeV) electromagnetically coupled WIMP favors $\Deln > 0$ and allows a sterile neutrino (\Deln~= 1, but not \Deln~= 2).  In contrast, a neutrino coupled light WIMP allows $\Deln = 0$ but disfavors $\Deln \gsim 1/2$ (at the fixed Planck $\Lambda$CDM value of \omb~shown).  The agreement between BBN and the CMB reinforces these conclusions as may be seen in the \Yp~-- \neff~plane in right hand panel of Fig.\,\ref{fig:ypvsdhneff} in terms of observed \Yp~and \neff.  Note that the CMB provides an independent, but currently very weak constraint on \Yp, allowing $\neff > 4$ at 95.5\% confidence.  The four solid (black) curves correspond to those in the left hand panel, and the dashed (blue) curve corresponds to the dashed curve in the left hand panel.  In agreement with the left panel, the combined constraints for the electromagnetically coupled case (to the left of the blue curve) slightly disfavor \Deln~= 0 and allow $\Deln > 1$ (but require $\Deln \lsim 3/2$).  For the neutrino coupled case (to the right of the blue curve), \Deln~= 0 is allowed and $\Deln = 1$ is disfavored (again, at the Planck $\Lambda$CDM \omb).  Thus, current BBN and CMB constraints challenge both SBBN (\Deln~= 0) and the presence of a sterile neutrino (\Deln~ = 1), whether or not there is a light WIMP.

\begin{figure}
\includegraphics[width=0.5\columnwidth]{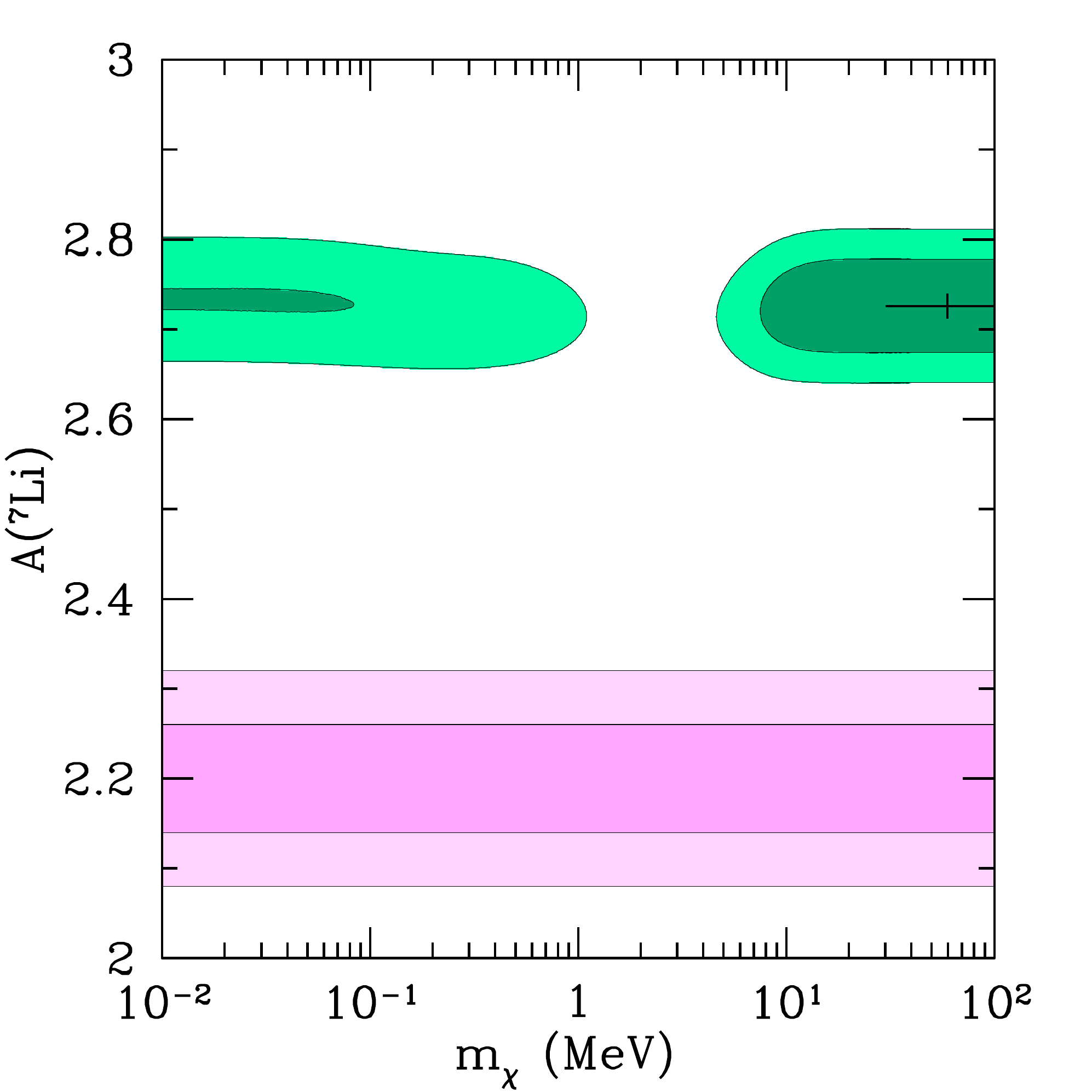}
  \caption{(Color online)  Predicted lithium abundance $A(\mathrm{Li}) \equiv 12+\log_{10}(\mathrm{Li/H})$, at the 68.3\%
and 95.5\% confidence allowed regions of our model parameters (upper curved green contours).  The lower, horizontal (pink-shaded) bands show the  primordial abundance inferred from halo stars in Ref.~\cite{spite}.  As in Fig.~\ref{fig:majorana-bbn-cmb}, the low-\mchi~contour corresponds to the unphysical $\Deln < 0$ regime.  The thickness of the contours in the vertical direction arises mainly from nuclear uncertainties on Li production.}
  \label{fig:lithium}
\end{figure}

A light WIMP does not help with the primordial lithium problem.  It is evident from Figs.~\ref{fig:bbncmb} and \ref{fig:ypvsdhneff}
that the measured and predicted D/H agree already without a light WIMP or equivalent neutrinos.   Fitting models to the deuterium abundance has the effect of fixing expansion timescales late in BBN, when deuterium is burned and $^7$Li (as $^7$Be) is created.  It was found for electromagnetically coupled WIMPs in Ref.~\cite{kngs1} that this guaranteed Li/H predictions close to the SBBN values.  This remains true for the neutrino coupled WIMPs considered here.  Moreover, the best fit for the neutrino coupled WIMP scenario is the high-\mchi~limit: the Li/H prediction of the best-fit model is identical to that with no light WIMP, and lower WIMP masses are only allowed with compensating values of \Deln, so that the Li/H prediction is the same for all \mchi.  This is illustrated in Fig.~\ref{fig:lithium}.

In summary, the BBN and CMB data are consistent without a light WIMP.  For a range of assumptions about the WIMP properties, the data imply lower limits to the allowed WIMP mass \mchi~in the MeV range.  An electromagnetically coupled WIMP slightly favors $\mchi \sim 8$ MeV, but the $\mchi \rightarrow \infty$ limit remains a quite good fit \cite{kngs1}.  The neutrino coupled WIMPs considered here do not allow for an even slightly better fit to the data than the $\mchi \rightarrow \infty$ limit, which is equivalent to no light WIMP at all.   The analysis here excludes all neutrino coupled WIMPs with masses below a few MeV, with specific limits varying from 4 to 9 MeV depending on the nature of the WIMP.  Unlike electromagnetically coupled WIMPs, there is no way to accommodate $\Deln = 1$ by lowering the mass of a neutrino coupled WIMP; the small accommodation toward $\Deln = 0$ allowed by a neutrino coupled WIMP is not strong enough to make $\Deln = 0$ a good fit to the data.   To avoid overinterpretation of data, it is important to consider models that provide fewer internal constraints than a ``minimal'' model with $\Deln = 0$ or with only equivalent neutrinos in addition to SM particles.  The particular implementations of such a nonminimal model considered here and in Ref.~\cite{kngs1} suggest that the $\Deln > 0$ found from abundance observations is robust to varying model assumptions.  

\acknowledgments

We are grateful to the Ohio State University Center for Cosmology and Astro-Particle Physics for support of G.\,S.'s research and for hosting K.\,M.\,N.'s visit during which most of the work described here was done.  We would also like to acknowledge useful comments and suggestions from John Beacom.   K.\,M.\,N. is pleased to acknowledge support from the Institute for Nuclear and Particle Physics at Ohio University, and at the University of South Carolina from U.S.~Department of Energy Award No.~DE-SC 0010 300 and Department of Energy Grant No.~DE-FG02-09ER41621.  G.\,S.\ also acknowledges the support and hospitality of the KITP at UCSB.  This research at the KITP was supported in part by the National Science Foundation under Grant No. NSF PHY11-25915.

\end{document}